\documentclass[aps,prd,onecolumn,superscriptaddress,nofootinbib,amsmath,amssymb,floats,floatfix,nofrontpages,showkeys,longbibliography]{revtex4}
\usepackage[dvips]{graphicx,color}
\usepackage{subfig}
\usepackage{times}
\usepackage{mathrsfs}  
\usepackage{xcolor}
 \usepackage{booktabs}
\usepackage[ colorlinks=true,
  urlcolor=blue,
  linkcolor=red,
  citecolor=blue
]{hyperref}

\RequirePackage{mathrsfs}
\RequirePackage{amsmath}
\RequirePackage{amssymb}
\RequirePackage{amsbsy}
\usepackage{graphicx}
\usepackage{multirow} 
\usepackage{caption}
\usepackage{tabularx}
\newcommand{\be}{\begin{equation}}
\newcommand{\ee}{\end{equation}}

\usepackage{color}
\usepackage{orcidlink}

\begin{document}
 
\title{Ringdown of a black hole sourced by a Burkert-density effective anisotropic source}

\author{Yi Yang
\orcidlink{0000-0003-1886-8716}}
\email{yiyang@mail.gufe.edu.cn}
\affiliation{School of Mathematics and Statistics, \\
Guizhou University of Finance and Economics, Guiyang, 550025, China}
\affiliation{School of Big Data Statistics, \\Guizhou University of Finance and Economics, Guiyang 550025, China}

\author{Gaetano Lambiase
\orcidlink{0000-0001-7574-2330}}
\email{lambiase@sa.infn.it}
\affiliation{Dipartimento di Fisica ``E.R Caianiello'', Universit degli Studi di Salerno, Via Giovanni Paolo II, 132 - 84084 Fisciano (SA), Italy}
\affiliation{Istituto Nazionale di Fisica Nucleare - Gruppo Collegato di Salerno - Sezione di Napoli, Via Giovanni Paolo II, 132 - 84084 Fisciano (SA), Italy}

\author{Ali \"Ovg\"un \orcidlink{0000-0002-9889-342X}}
\email{ali.ovgun@emu.edu.tr}
\affiliation{Physics Department, Eastern Mediterranean University, Famagusta, 99628 North Cyprus via Mersin 10, Turkiye}

\author{Dong Liu}
\email{dongliuvv@yeah.net}
\affiliation{College of Physics, Guizhou University, Guiyang, 550025, China}

\author{Zheng-Wen Long}
\email{zwlong@gzu.edu.cn}
\affiliation{College of Physics, Guizhou University, Guiyang, 550025, China}

\begin{abstract}
We construct a static, spherically symmetric black hole spacetime sourced by an effective anisotropic matter distribution whose energy density follows the cored Burkert profile. 
The source should be understood as a Burkert-density effective fluid, rather than as a microscopic model of pressureless collisionless dark matter. 
Solving the Einstein equations under this closure condition, we obtain an analytic Schwarzschild-like metric that reduces smoothly to the vacuum Schwarzschild solution when the halo contribution vanishes. 
We then study axial gravitational perturbations of this geometry and determine the associated quasinormal spectrum using complementary frequency-domain and time-domain methods. 
We find that increasing either the core radius \(r_0\) or the central density \(\rho_0\) shifts the ringdown toward lower frequency and weaker damping, with the effect of \(r_0\) being more pronounced. 
The close agreement among the numerical extractions supports the reliability of the results. 
Our analysis provides a useful benchmark for assessing how a cored Burkert-type effective environment can modify black hole ringdown.
\end{abstract}


\maketitle
\flushbottom

\section{Introduction}\label{sec:intro}

Dark matter (DM) constitutes about $25\%$ of the total energy density of the Universe and plays a central role in the formation and evolution of cosmic structure \cite{Planck2018}. Its presence is inferred from a wide range of observations, including galaxy rotation curves, strong and weak lensing, satellite kinematics, and the cosmic microwave background, all of which require a matter component beyond the baryonic sector \cite{deBlok2010,Donato2009,Seo:2020chg}. On galactic scales, however, the inner structure of DM halos remains under debate. Standard cold-DM simulations typically predict cuspy density profiles, such as the Navarro--Frenk--White (NFW) profile, whereas observations of dwarf and low-surface-brightness galaxies often favor cored profiles with nearly constant central density. Among these, the Burkert profile has proved particularly successful in fitting observed rotation curves over a broad radial range \cite{Navarro1996,Burkert:1995yz,SalucciBurkert2000,Donato2009}.

Black holes embedded in DM halos provide a useful setting in which to study how an extended matter distribution modifies strong-field geometry and how such modifications may appear in observable quantities. Several related directions have recently been explored. An Einstein cluster of collisionless particles was shown to generate a central DM spike with potentially relevant consequences for galactic centers \cite{Maeda:2024tsg}. A vector-field model reproducing Einstein-cluster behavior was later proposed, thereby providing a relativistic framework for equilibrium and stability analyses \cite{Fernandes:2025lon}. The optical appearance and gravitational wave signatures of black holes in generic DM environments were investigated in \cite{Figueiredo:2023gas}, where profile-dependent changes in shadows and ringdown spectra were identified. Black hole spacetimes surrounded by specific halo models, including the Dekel--Zhao and pseudo-isothermal profiles, have also been constructed and analyzed from the perspective of geodesics, lensing, and quasinormal modes \cite{Ovgun:2025bol,Yang:2023tip}. Further studies have considered wave or fuzzy DM \cite{Pantig:2022sjb}, superradiant and ringdown effects in rotating backgrounds \cite{Liu:2022ygf}, weak deflection in different DM scenarios \cite{Pantig:2022toh}, and wormhole geometries supported by galactic DM distributions \cite{Islam:2018ciy}.

In phenomenological applications, halo models are often grouped into cuspy and cored classes. While cuspy profiles are a natural outcome of collisionless cold-DM simulations, cored profiles are often preferred by observations at small galactic radii. A widely used cored model is the \emph{Burkert} profile \cite{Burkert:1995yz},
\begin{equation} \label{Burkertprofile}
\rho_{\rm Bur}(r)=\frac{\rho_0}{\left(1+r/r_0\right)\left[1+(r/r_0)^2\right]},
\end{equation}
where $\rho_0$ and $r_0$ denote the central density and core radius, respectively. This profile reproduces both the inner flattening of the density and the outer behavior of observed rotation curves \cite{Burkert:1995yz,SalucciBurkert2000}, and the parameters $(\rho_0,r_0)$ are known to obey empirical scaling relations across galaxy samples \cite{Donato2009}.

It is important to distinguish the density profile from the full relativistic matter model. 
The Burkert profile fixes only the radial dependence of the energy density. 
It does not determine the pressure components of the stress-energy tensor. 
In particle dark matter models around black holes, the central object can strongly reshape the inner distribution and may generate a spike or an inner depletion region \cite{Gondolo:1999ef,Sadeghian:2013laa,Speeney:2022}. 
In Einstein-cluster-type constructions, the halo is supported by tangential motion and the radial pressure vanishes \cite{Cardoso:2021wlq,Shen:2024}. 
Recent analytic models also show that an inner edge of the halo can be important for satisfying the energy conditions \cite{Shen:2024b}. 
The present work follows a different route. 
We use the Burkert profile as an effective density input and close the Einstein equations with an anisotropic-fluid condition. 
Therefore, the resulting source should be understood as a Burkert-density effective anisotropic fluid, rather than as a microscopic model of pressureless collisionless dark matter.

With this interpretation in mind, embedding a black hole in such an effective Burkert-type environment raises two natural questions. First, how does the halo stress--energy modify the background geometry of a static, spherically symmetric black hole? Second, how do these geometric corrections affect observables such as quasinormal modes, ringdown signals, and possibly lensing or shadow features? A practical way to address this problem is to begin with a halo geometry constrained by the observed circular velocity and then solve the Einstein equations with an effective matter stress--energy tensor while requiring that the central object reduce to Schwarzschild in the appropriate limit \cite{Matos2000,Xu:2018wow}. This construction yields an analytic metric in which the influence of the halo parameters on the effective perturbation potential can be studied explicitly.
The study of black hole ringing traces back to the foundational analyses of perturbations of Schwarzschild spacetime by Regge and Wheeler and by Zerilli, together with the early recognition that black holes respond to disturbances through damped oscillations and radiative relaxation, as shown in classic works on gravitational radiation damping, scattering, infall, and rotating black hole perturbations by Thorne, Vishveshwara, Press, Davis \textit{et al.}, Teukolsky, Detweiler, and Thorne \cite{Regge:1957td,Zerilli:1970wzz,Thorne:1968zz,Vishveshwara:1970zz,Press:1971wr,Davis:1971gg,Teukolsky:1972my,Detweiler:1980uk,Thorne:1997cw}. In the era of gravitational wave astronomy inaugurated by the first direct observation of a binary-black hole merger \cite{LIGOScientific:2016aoc,LIGOScientific:2016aoc}, quasinormal modes (QNMs) have become central to black hole spectroscopy, tests of strong-field gravity, and the broader program linking compact objects, gravitational waves, and fundamental physics \cite{Kokkotas:1999bd,Nollert:1999ji,Berti:2009kk,Konoplya:2011qq,Berti:2015itd,Cardoso:2016rao,Barack:2018yly}.

On the computational side, a remarkably rich toolkit has been developed for extracting QNM spectra. This includes the semianalytic WKB framework introduced by Schutz and Will and extended to higher orders by Iyer and Will, along with later refinements and Pad\'e-improved implementations that substantially enhance accuracy for barrier-penetration problems \cite{Schutz:1985km,Iyer:1986np,Konoplya:2003ii,Konoplya:2019hlu,Matyjasek:2019eeu}. These methods are complemented by phase-integral and time-domain approaches, analyses of asymptotic spectra, studies in de Sitter and hairy backgrounds, modern Borel-based resummation techniques, and recent developments addressing mode stability and machine-learning-assisted solutions of black hole perturbation equations \cite{Andersson:1995vi,Andersson:1996cm,Andersson:2003fh,Zhidenko:2003wq,Zhidenko:2005mv,Daghigh:2008jz,Daghigh:2011ty,Hatsuda:2019eoj,Eniceicu:2019npi,Berti:2022xfj,Luna:2022rql,Lepe:2004kv}. In the eikonal regime, the spectrum often admits an appealing geometric interpretation in terms of unstable photon orbits and related optical observables, as discussed for Kerr, Kerr--Newman, and shadow/QNM correspondences; however, this link is not universal and can fail in more general settings, where mode competition and spectral structure become richer \cite{Yang:2012he,Li:2021zct,Yang:2021zqy,Feng:2022otu,Gogoi:2024vcx,Konoplya:2017wot,Glampedakis:2019dqh,Churilova:2019jqx,Dias:2022oqm}.

At the same time, QNMs are closely intertwined with wave scattering, absorption, greybody factors, and superradiant phenomena. This broader perspective is reflected in classic and modern studies of electromagnetic and scalar absorption by black holes, black hole greybody factors, and the joint analysis of QNMs with transmission properties in a variety of settings \cite{Fabbri:1975sa,Unruh:1976fm,Klebanov:1997cx,Benone:2014qaa,Macedo:2014uga,Crispino:2015gua,Leite:2017zyb,Rincon:2018ktz,Rincon:2020cos,Yang:2022ifo,Gogoi:2023lvw,Pantig:2022gih,Okyay:2021nnh}. This line of work has helped clarify how the ringdown spectrum responds not only to the underlying geometry but also to the effective scattering potential generated by matter sources, modified interactions, or additional fields. More recently, QNMs have been investigated across a wide variety of nonvacuum and beyond-GR spacetimes \cite{Panotopoulos:2019qjk,Panotopoulos:2020mii,Rincon:2021gwd,Gonzalez:2021vwp,Gonzalez:2022ote,Chabab:2016cem,Chabab:2017knz,Boudet:2022wmb,Aoki:2020iwm,Zhang:2020khz,Kruglov:2021qzd,Luo:2022gdz,Breton:2021mju,Zhao:2023uam,Anacleto:2021qoe,Lambiase:2023hng,sekhmani_electromagnetic_2023,Parbin:2022iwt,karmakar_quasinormal_2022,Bora:2022qwe,Gogoi:2022wyv,Gogoi:2021cbp,Gogoi:2021mhi,Gogoi:2021dkr,Gogoi:2020ypn,Gogoi:2019zaz}. In parallel, increasing attention has been devoted to the interplay between ringdown and imaging observables, including black hole shadows, photon rings, lensing rings, and superradiant shadow evolution, further emphasizing the diagnostic power of combining gravitational wave and electromagnetic probes \cite{Gralla:2019xty,Wang:2019skw,Chen:2022nbb,Roy:2021uye,Jafarzade:2020ova}. Against this broader backdrop, it is timely to investigate how realistic matter environments alter the ringdown spectrum of astrophysical black holes; in particular, black holes embedded in dark matter halos provide a natural setting in which environmental effects can be studied quantitatively and connected to strong-gravity observations.

The motivation for the present work is twofold. 
From the astrophysical perspective, the Burkert profile is well supported by galactic rotation-curve observations as a cored dark-matter density profile \cite{Burkert:1995yz,SalucciBurkert2000,Donato2009}. This makes it a useful density input for constructing an effective matter environment around black holes. From the gravitational perspective, the ringdown phase is one of the cleanest probes of the near-horizon and photon-sphere geometry, and thus offers a natural arena in which to search for environmental imprints on strong-field observables \cite{Berti:2009kk,Konoplya:2011qq,CardosoPani2019,Figueiredo:2023gas}. Although black holes embedded in generic or model-dependent dark matter halos have been investigated in a variety of settings, comparatively little is known about how a Burkert-density effective halo modifies the gravitational perturbation spectrum of the central compact object in a fully relativistic framework \cite{Xu:2018wow,Yang:2023tip,Ovgun:2025bol}. This issue is particularly important because cored and cuspy profiles can induce qualitatively different deformations of the effective potential and may therefore leave distinguishable signatures in the late-time ringdown signal.

In this sense, black holes surrounded by matter provide an important class of ``dirty'' spacetimes in which environmental modifications of the ringdown spectrum can be studied quantitatively \cite{Berti:2009kk,Konoplya:2011qq}. The aim of this paper is to provide a concrete benchmark for such effects in the case of a Burkert-density effective halo. To this end, we construct a static, spherically symmetric black hole spacetime sourced by the Burkert profile and examine its basic geometric and physical properties. We then study axial gravitational perturbations of this background and compute the corresponding quasinormal spectrum using complementary frequency domain and time-domain methods. In this way, we isolate how the halo parameters, namely the central density $\rho_0$ and core radius $r_0$, modify the effective potential, the oscillation frequencies, and the damping rates of the ringdown modes. Our goal is therefore not only to clarify the role of a dark-matter-density-inspired effective environment in black hole spectroscopy, but also to provide a useful reference framework for future studies of environmental effects in gravitational wave phenomenology.

\section{Spherically symmetric black hole sourced by a Burkert-density effective anisotropic fluid}\label{PIandF}

We now construct a static, spherically symmetric black hole solution sourced by a Burkert-density effective matter distribution. We begin with the line element, in units $G=c=1$,
\begin{equation}
ds^2=-\mathcal{P}(r)\,dt^2+\frac{dr^2}{\mathcal{Q}(r)}+r^2 d\Omega^2,
\qquad d\Omega^2=d\theta^2+\sin^2\theta\,d\phi^2,
\label{eq:metric_PQ}
\end{equation}
and model the matter source as an anisotropic fluid with stress--energy tensor
\begin{equation}
T^{\mu}{}_{\nu}=\mathrm{diag}\bigl(-\rho(r),\,P_r(r),\,P_t(r),\,P_t(r)\bigr),
\label{eq:SET_aniso}
\end{equation}
where $\rho$ is the energy density and $P_r$ and $P_t$ are the radial and tangential pressures. Substituting Eq.~\eqref{eq:metric_PQ} into the Einstein equations, $G^{\mu}{}_{\nu}=8\pi T^{\mu}{}_{\nu}$, gives three independent ordinary differential equations, with primes denoting derivatives with respect to $r$:
\begin{align}
&\frac{1}{r^2}\Big(\mathcal{Q}-1+r\,\mathcal{Q}'\Big)
=8\pi T^{t}{}_{t}
=-8\pi\,\rho(r),
\label{eq:Ein_tt_num}\\[3pt]
&\frac{\mathcal{Q}-1}{r^2}+\frac{\mathcal{Q}}{\mathcal{P}}\frac{\mathcal{P}'}{r}
=8\pi T^{r}{}_{r}
=8\pi\,P_r(r),
\label{eq:Ein_rr_num}\\[3pt]
&\frac{\mathcal{Q}}{2\mathcal{P}}\,\mathcal{P}''
-\frac{\mathcal{Q}}{4\mathcal{P}^2}\,(\mathcal{P}')^2
+\frac{\mathcal{Q}'}{4\mathcal{P}}\,\mathcal{P}'
+\frac{\mathcal{Q}'}{2r}
+\frac{\mathcal{Q}}{2r\mathcal{P}}\,\mathcal{P}'
=8\pi T^{\theta}{}_{\theta}
=8\pi\,P_t(r),
\label{eq:Ein_thth_num}\\[3pt]
&8\pi T^{\theta}{}_{\theta}
=8\pi T^{\phi}{}_{\phi}
=8\pi\,P_t(r).
\label{eq:Ein_phph_num}
\end{align}

Following the standard treatment of spherically symmetric spacetimes, we introduce the mass function $\mathfrak{m}(r)$ through
\begin{equation}
\mathcal{Q}(r)=1-\frac{2\mathfrak{m}(r)}{r},
\label{eq:mass_def}
\end{equation}
from which one immediately obtains
\begin{equation}
\mathfrak{m}'(r)=4\pi r^2 \rho(r).
\label{eq:mprime}
\end{equation}
For a given density profile, $\mathfrak{m}(r)$ represents the Misner--Sharp mass enclosed within radius $r$. We decompose it as
\begin{equation}
\mathfrak{m}(r)=M+\mathfrak{m}_h(r),\qquad
\mathfrak{m}_h(r)=4\pi\int_{0}^{r}\rho(r')\,r'^2\,dr',
\label{eq:m_split}
\end{equation}
where $M$ is the central black hole mass and $\mathfrak{m}_h(r)$ is the contribution from the surrounding matter distribution.

We now adopt the Burkert density profile given in Eq. \eqref{Burkertprofile} and substitute it into Eq.~\eqref{eq:m_split}. The integral can be evaluated in closed form, yielding
\begin{equation}
\mathfrak{m}_h(r)=\pi\rho_0 r_0^3\left[
2\ln\!\left(1+\frac{r}{r_0}\right)
+\ln\!\left(1+\frac{r^2}{r_0^2}\right)
-2\arctan\!\left(\frac{r}{r_0}\right)
\right].
\label{eq:mh_burkert}
\end{equation}

The integral in Eq.~(10) should be understood within the effective construction adopted here. 
It gives a smooth analytic continuation of the Burkert density profile into the strong-field region, rather than a microscopic collisionless halo filling the whole spacetime without an inner edge. 
A more realistic particle-dark-matter halo with an inner cutoff \(r_{\rm in}\) would require a different piecewise construction and suitable matching conditions.

For the perturbation analysis, it is useful to estimate the relative size of the effective matter contribution. 
We introduce the dimensionless ratio
\[
\eta(r)=\frac{m_h(r)}{M},
\]
which measures the effective matter mass enclosed within radius \(r\) relative to the central black-hole mass. 
Since the ringdown is mainly controlled by the geometry near the photon sphere, \(\eta(r_c)\) provides a useful diagnostic of the strength of the environmental correction. 

For the benchmark choice \(M=1\), \(\rho_0=0.1\), and \(r_0=0.1\), we find
\[
r_h\simeq 2.00569,\qquad r_c\simeq 3.00878,
\]
and hence
\[
\eta(r_h)\simeq 2.84\times10^{-3},\qquad
\eta(r_c)\simeq 3.33\times10^{-3}.
\]
Moreover, for the density scan, when \(M=1\), \(r_0=0.1\), and \(\rho_0\) is varied from \(0.1\) to \(1.0\), we find that \(\eta(r_c)\) increases monotonically from \(3.33\times10^{-3}\) to \(3.36\times10^{-2}\). 
Thus, in this density scan the effective matter contribution near the photon sphere stays below \(3.4\%\), indicating a weak effective-environment regime.
For the core-radius scan with \(M=1\), \(\rho_0=0.1\), and \(r_0\) varied from \(0.1\) to \(1.0\), \(\eta(r_c)\) is more sensitive to \(r_0\). 
It reaches \(2.04\times10^{-1}\) at \(r_0=0.5\) and increases further for larger \(r_0\), indicating that the largest-\(r_0\) cases no longer belong to the weak effective-environment regime. 
Therefore, the small-\(r_0\) cases correspond to a weak effective environment, whereas the larger-\(r_0\) cases should be regarded as a phenomenological exploration of stronger effective environmental corrections.
If one wants to define a global mass for a finite halo, an outer cutoff \(R_h\) should be introduced. 
The corresponding total mass is then
\[
M_{\rm tot}(R_h)=M+m_h(R_h).
\]
This global quantity is not directly used in the QNM calculation, which is mainly governed by the local strong-field contribution near the photon sphere.

It then follows from Eq.~\eqref{eq:mass_def} that
\begin{equation}
\mathcal{Q}(r)=1-\frac{2M}{r}
-\frac{2\pi\rho_0 r_0^3}{r}\left[
2\ln\!\left(1+\frac{r}{r_0}\right)
+\ln\!\left(1+\frac{r^2}{r_0^2}\right)
-2\arctan\!\left(\frac{r}{r_0}\right)
\right].
\label{eq:Q_burkert}
\end{equation}

To close the system and obtain an explicitly self-consistent effective model, 
we further impose the anisotropic-fluid condition~\cite{Bolokhov:2025zva}
\begin{equation}
P_r(r)=-\rho(r),\qquad 
P_t(r)=-\rho(r)-\frac{r}{2}\rho'(r).
\label{eq:closure_A_compact}
\end{equation}
This condition should be understood as an effective closure relation. The Burkert profile fixes only the energy density and does not by itself determine the pressure sector of the relativistic stress-energy tensor. 
In Einstein-cluster descriptions of dark matter halos, the radial pressure is zero and the tangential pressure is generated by orbital motion of collisionless particles. 
In such models, the halo may also possess an inner edge outside the black-hole horizon. 
By contrast, the present model is a smooth effective extension of the Burkert density profile into the strong-field region. 
Therefore, the lower bound on the inner halo radius obtained in Einstein-cluster models does not directly apply to the present effective source, because the underlying stress-energy tensor is different.
For this reason, throughout the rest of this work we interpret the solution as a black hole sourced by a Burkert-density effective anisotropic fluid. 

Under this effective closure condition, one finds $\mathcal{P}(r)=\mathcal{Q}(r)\equiv \mathcal{F}(r)$. 
The metric therefore takes the form
\begin{equation}
ds^2=-\mathcal{F}(r)\,dt^2+\frac{dr^2}{\mathcal{F}(r)}+r^2 d\Omega^2,
\label{eq:final_metric_Z}
\end{equation}
with
\begin{equation}
\mathcal{F}(r)=1-\frac{2M}{r}
-\frac{2\pi\rho_0 r_0^3}{r}\left[
2\ln\!\left(1+\frac{r}{r_0}\right)
+\ln\!\left(1+\frac{r^2}{r_0^2}\right)
-2\arctan\!\left(\frac{r}{r_0}\right)
\right].
\label{eq:Z_burkert_final}
\end{equation}

The analytic spacetime defined by Eqs.~\eqref{eq:final_metric_Z} and \eqref{eq:Z_burkert_final} represents a black hole of mass \(M\) sourced by a Burkert-density effective anisotropic matter distribution.
Near the center, setting $x=r/r_0\ll 1$ gives
\begin{equation}
\mathcal{F}(r)=1-\frac{2M}{r}-\frac{8\pi\rho_0}{3}\,r^2+O(r^4),
\end{equation}
so that, for $M>0$, the spacetime retains the usual Schwarzschild-type central singularity. In the asymptotic region, one has $\mathcal{F}(r)\to 1$ because
$m_h(r)/r\to 0$. Thus the metric coefficients approach their
Minkowski values. However, since the Burkert density falls as
$\rho\sim r^{-3}$, the effective mass grows logarithmically
at very large radius. Consequently, a strictly finite ADM mass requires
an outer cutoff $R_h$, or equivalently a matching to an exterior vacuum
region. In the present work we use the untruncated analytic profile only
as a local effective description of the strong-field region relevant for
the photon sphere and ringdown.

\begin{figure}[htbp]
\vspace{15pt}
\begin{center}
\includegraphics[scale=0.4]{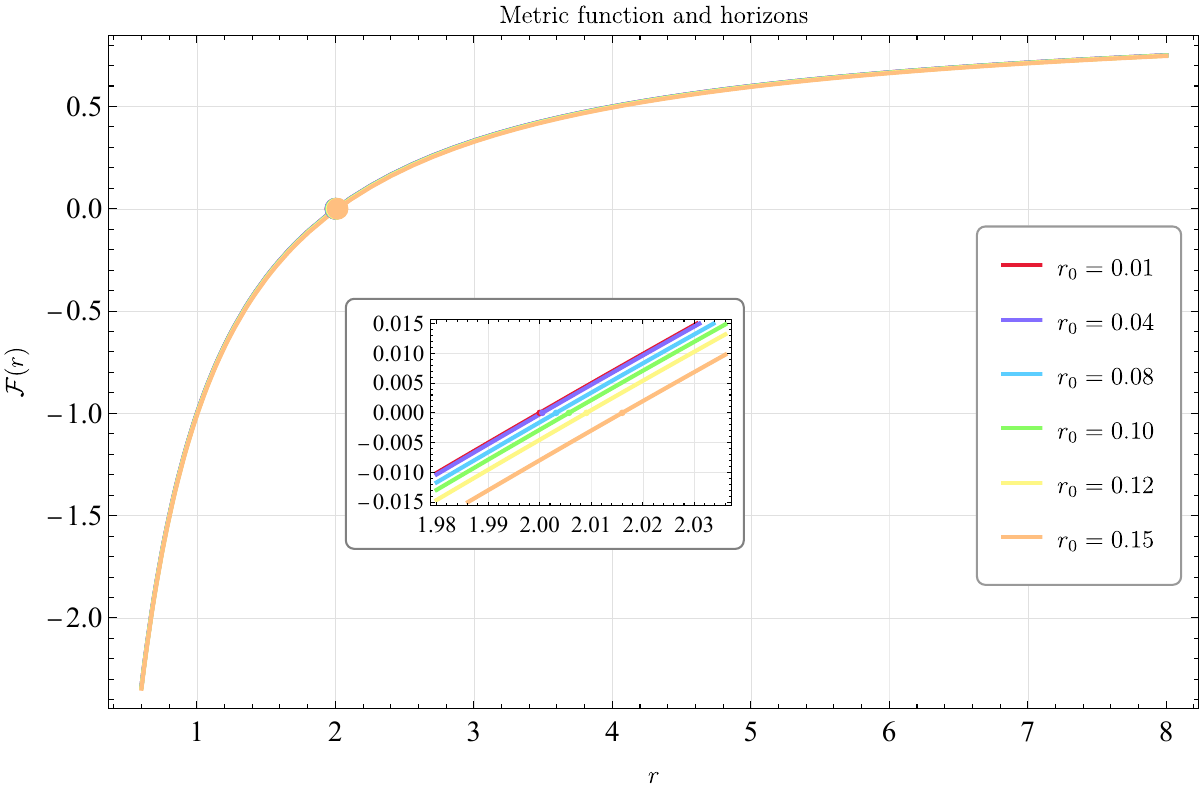}\qquad
\includegraphics[scale=0.4]{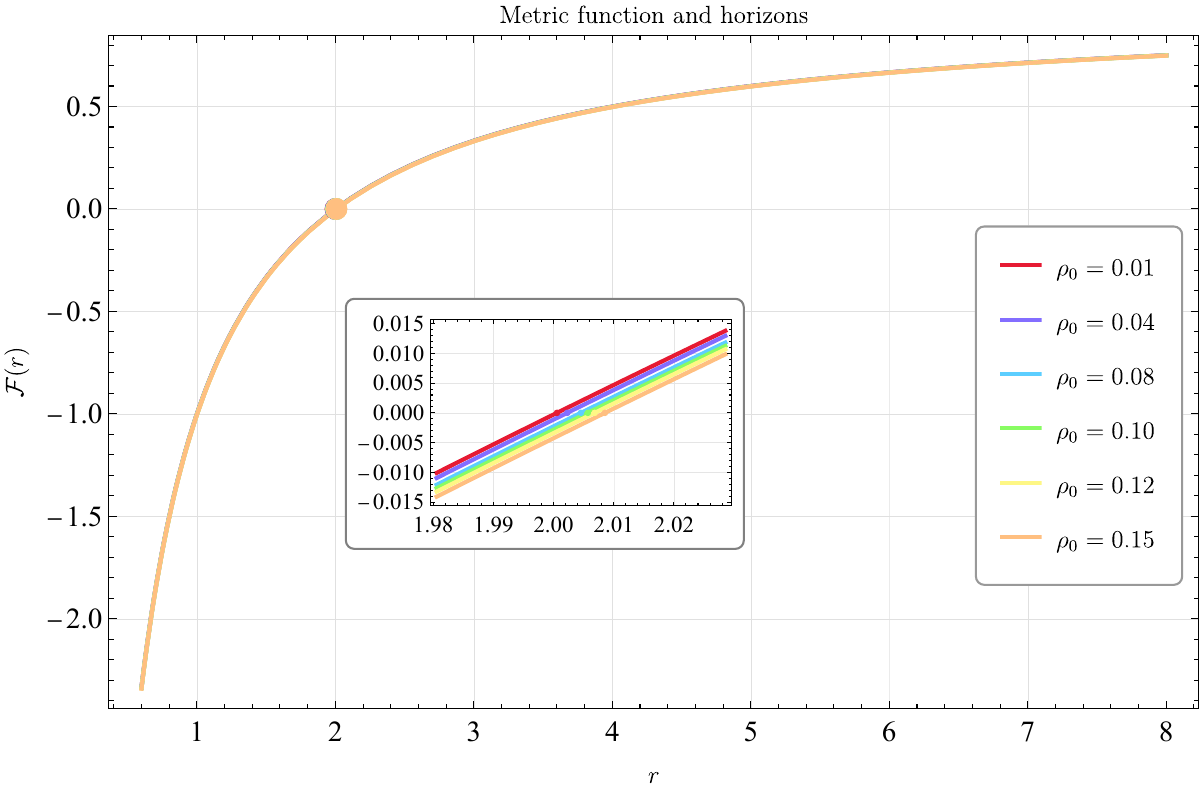}
\end{center}
\setlength{\abovecaptionskip}{-0.1cm}
\setlength{\belowcaptionskip}{0.8cm}
\caption{Metric function and horizon location for the Burkert black hole.
In each panel, we plot the metric function $\mathcal{F}(r)$ for several parameter choices, and the event horizon is identified by the unique positive root $\mathcal{F}(r_h)=0$ (marked by dots).
Left panel: we fix $M=1$ and $\rho_0=0.1$ and vary $r_0\in\{0.01,0.04,0.08,0.10,0.12,0.15\}$.
Right panel: we fix $M=1$ and $r_0=0.1$ and vary $\rho_0\in\{0.01,0.04,0.08,0.10,0.12,0.15\}$.
 In both cases, the horizon lies close to the Schwarzschild value $r_h=2M=2$ and shifts slightly outward as $r_0$ or $\rho_0$ increases.}
\label{fig:horizon_two_panels}
\end{figure}

Fig.~\ref{fig:horizon_two_panels} displays the metric function $\mathcal{F}(r)$ for the Burkert black hole. The event horizon is identified as the largest positive root of $\mathcal{F}(r_h)=0$. In the left panel, we fix $M=1$ and $\rho_0=0.1$ and vary the core radius over $r_0\in\{0.01,0.04,0.08,0.10,0.12,0.15\}$. In the right panel, we fix $M=1$ and $r_0=0.1$ and vary the central density over $\rho_0\in\{0.01,0.04,0.08,0.10,0.12,0.15\}$. In each case, the horizon is marked by the intersection of the corresponding curve with the horizontal axis. The insets enlarge the near-horizon region and show that the outer horizon remains unique. As expected, the horizon stays close to the Schwarzschild value $r_h=2M=2$, with only a small shift induced by the halo parameters.

To further verify the asymptotic behavior of the spacetime, we calculate several curvature invariants associated with the metric~\eqref{eq:Z_burkert_final}. The Ricci scalar is given by
\begin{equation}
R=
\frac{
8\pi\rho_0 r_0^3
\left(
r^3+2r^2r_0+3rr_0^2+4r_0^3
\right)
}
{(r+r_0)^2(r^2+r_0^2)^2},
\end{equation}
while the Ricci-tensor contraction takes the form
\begin{equation}
R_{\mu\nu}R^{\mu\nu}
=
128\pi^2\rho_0^2
\left[
\frac{r_0^6}
{(r+r_0)^2(r^2+r_0^2)^2}
+
\frac{
r_0^6
\left(
r^3-r r_0^2-2r_0^3
\right)^2
}
{4(r+r_0)^4(r^2+r_0^2)^4}
\right].
\end{equation}
We also calculate the Kretschmann scalar
\begin{equation}
K=\frac{1}{r^6}
\left\{
\left[
-A r^2\mathcal{H}''(r)
+2Ar\mathcal{H}'(r)
-2\left(2M+A\mathcal{H}(r)\right)
\right]^2
+
4\left[
-Ar\mathcal{H}'(r)
+2M+A\mathcal{H}(r)
\right]^2
+
4\left[
2M+A\mathcal{H}(r)
\right]^2
\right\},
\end{equation}
where
\begin{align}
\mathcal{H}(r)
&=
2\ln\left(1+\frac{r}{r_0}\right)
+\ln\left(1+\frac{r^2}{r_0^2}\right)
-2\arctan\left(\frac{r}{r_0}\right), \nonumber\\
\mathcal{H}'(r)
&=
\frac{4r^2}
{(r+r_0)(r^2+r_0^2)}, \nonumber\\
\mathcal{H}''(r)
&=
-\frac{
4r\left(r^3-rr_0^2-2r_0^3\right)
}
{(r+r_0)^2(r^2+r_0^2)^2}\\
A&=2\pi\rho_0 r_0^3, \nonumber.
\label{Hdefs}
\end{align}

Using these exact expressions obtained from the metric, we find
\begin{equation}
\lim_{r\rightarrow\infty}R=0,
\qquad
\lim_{r\rightarrow\infty}
R_{\mu\nu}R^{\mu\nu}=0,
\qquad
\lim_{r\rightarrow\infty}K=0.
\end{equation}
Together with
\begin{equation}
\lim_{r\rightarrow\infty}\mathcal{F}(r)=1,
\end{equation}
these results confirm that the spacetime is asymptotically flat. 

\subsection{Energy conditions}
The energy conditions provide a useful consistency check for the effective anisotropic source. 
However, they are not sufficient to identify the source with microscopic collisionless dark matter. 
They only test whether the effective stress-energy tensor behaves reasonably in the exterior region relevant to the perturbation problem. 
With this limitation in mind, we now examine the weak, dominant, and strong energy conditions.

For the anisotropic fluid defined in Eq.~\eqref{eq:SET_aniso}, the standard energy conditions are
\begin{align}
\text{WEC:}\quad & \rho\ge 0,\qquad \rho+P_r\ge 0,\qquad \rho+P_t\ge 0, \label{eq:WEC_def}\\
\text{DEC:}\quad & \rho\ge |P_r|,\qquad \rho\ge |P_t|, \label{eq:DEC_def}\\
\text{SEC:}\quad & \rho+P_r\ge 0,\qquad \rho+P_t\ge 0,\qquad \rho+P_r+2P_t\ge 0. \label{eq:SEC_def}
\end{align}
Using the pressure relations given in Eq.~\eqref{eq:closure_A_compact}, one immediately obtains
\begin{align}
\rho+P_r &= 0, \label{eq:rhoPr_saturate}\\
\rho+P_t &= -\frac{r}{2}\rho'(r), \label{eq:rhoPt_key}\\
\rho-|P_r| &= 0. \label{eq:DEC_Pr_saturate}
\end{align}
For the Burkert profile given in Eq. \eqref{Burkertprofile} with $x\equiv \frac{r}{r_0}$,
one finds the explicit tangential pressure
\begin{equation}
P_t(r)=\rho_0\,\frac{x^3-x-2}{2(1+x)^2(1+x^2)^2}.
\end{equation}

Since $\rho_0\ge 0$ and $\rho'(r)\le 0$ for $r>0$, Eq.~\eqref{eq:rhoPt_key} immediately implies $\rho+P_t\ge 0$. Therefore, the weak energy condition is satisfied, while the relation $\rho+P_r=0$ shows that the radial part saturates the bound.

\begin{figure}[b]
\vspace{25pt}
\begin{center}
\includegraphics[scale=0.4]{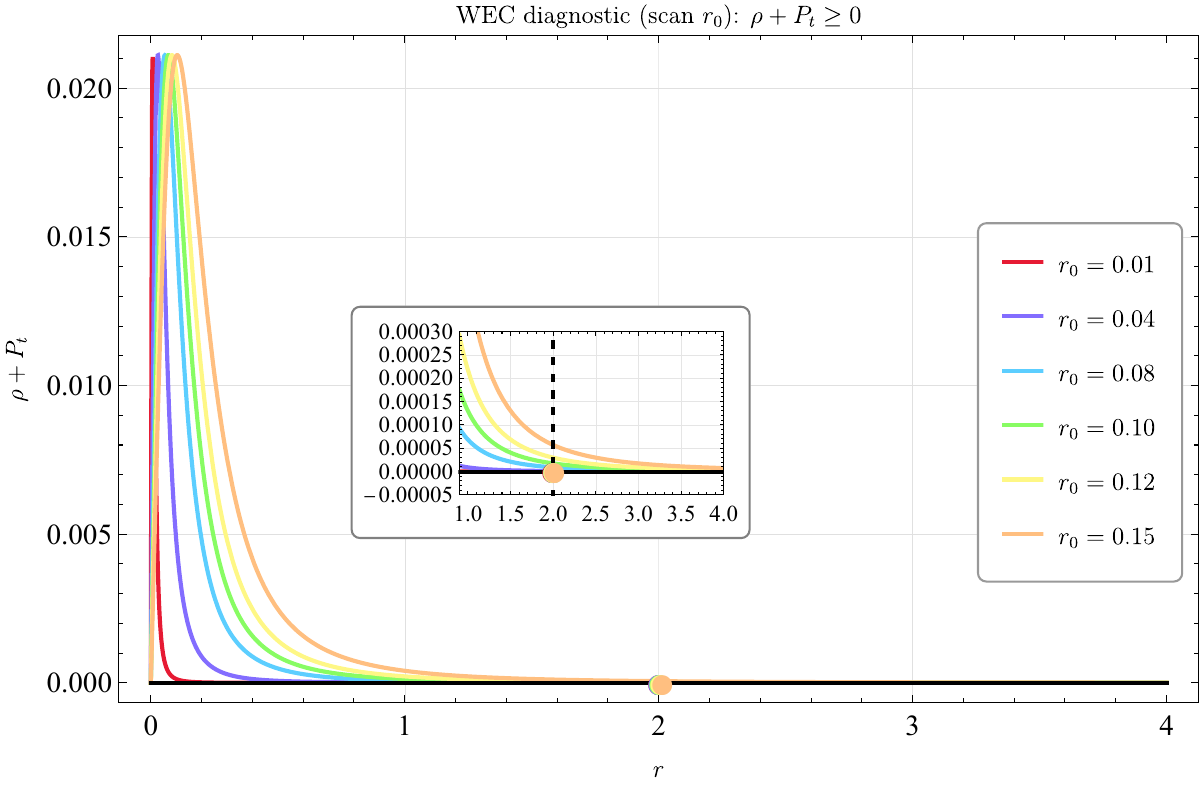}\qquad
\includegraphics[scale=0.4]{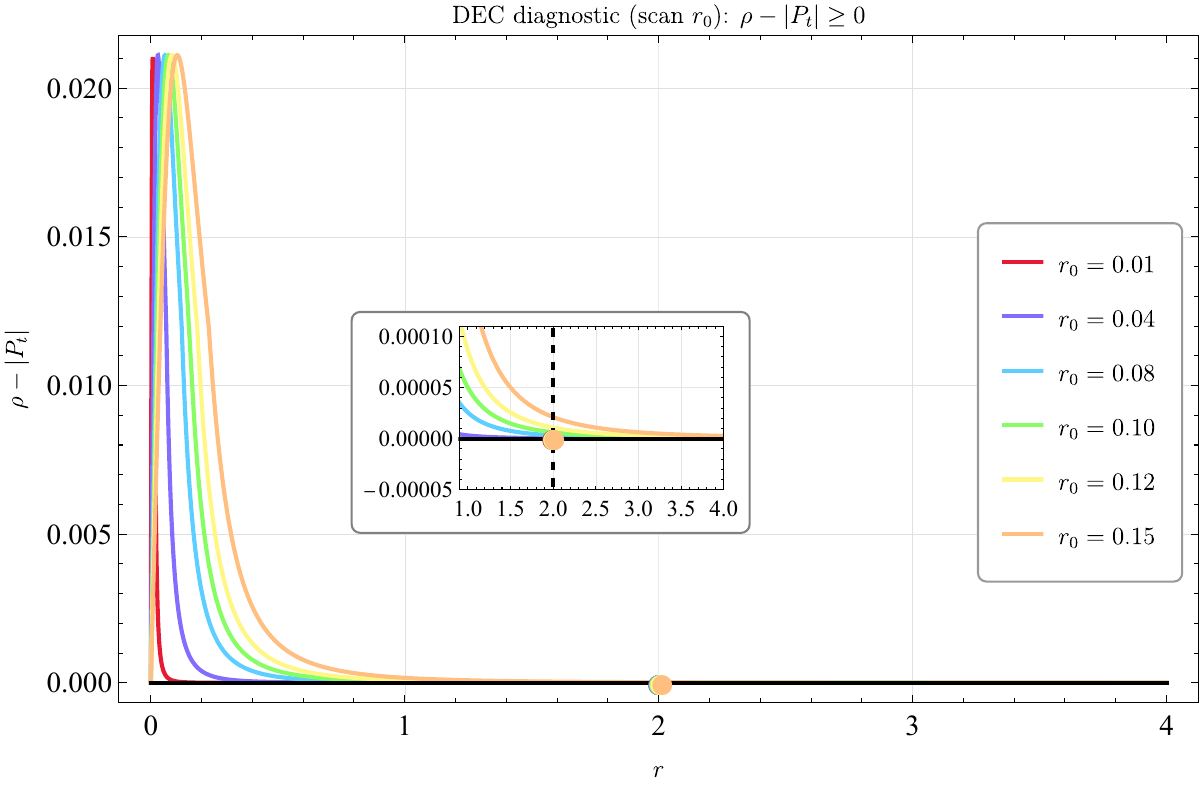}
\includegraphics[scale=0.4]{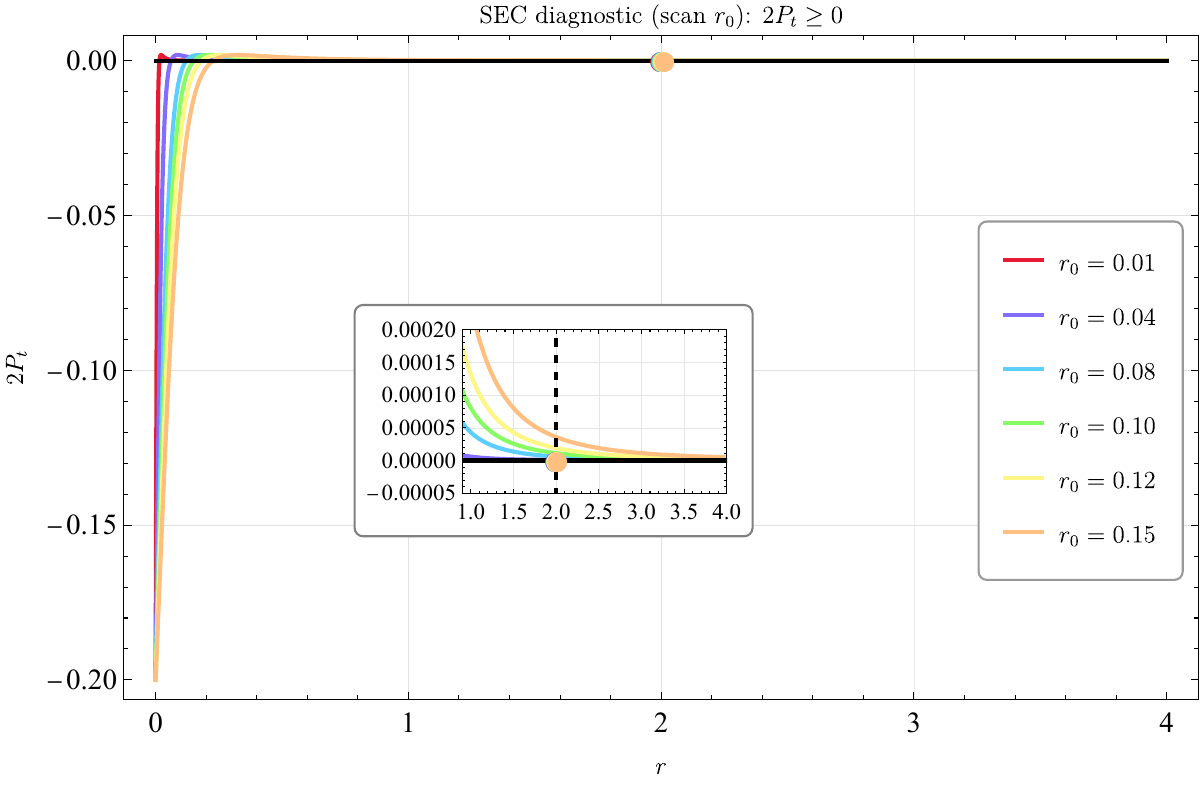}
\end{center}
\setlength{\abovecaptionskip}{-0.1cm}
\setlength{\belowcaptionskip}{-0.8cm}
\caption{Energy-condition diagnostics for the Burkert model in a scan over $r_0$.
We fix $M=1$ and $\rho_0=0.1$, and vary $r_0\in\{0.01,0.04,0.08,0.10,0.12,0.15\}$.
The plotted quantities are $\rho+P_t$ (WEC), $\rho-|P_t|$ (DEC), and $2P_t=\rho+P_r+2P_t$ (SEC).
The horizontal line denotes the zero baseline separating the allowed ($\ge 0$) and disallowed ($<0$) regions.
Colored dots indicate the horizon positions $r_h$ obtained from $\mathcal{F}(r_h)=0$.
The inset enlarges the near-horizon region and shows explicitly that the energy conditions are satisfied in the exterior domain relevant to the perturbation analysis.}
\label{fig:ec_scan_r0}
\end{figure}

\begin{figure}[t!]
\vspace{25pt}
\begin{center}
\includegraphics[scale=0.4]{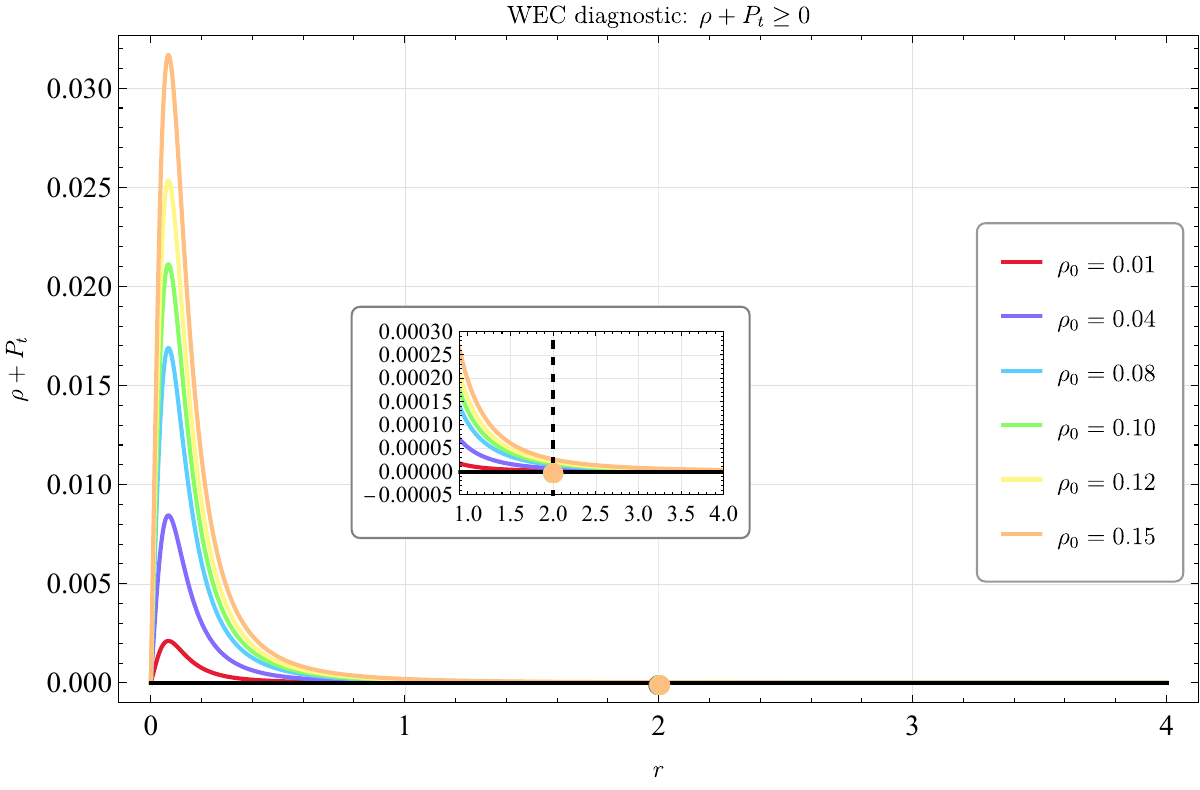}\qquad
\includegraphics[scale=0.4]{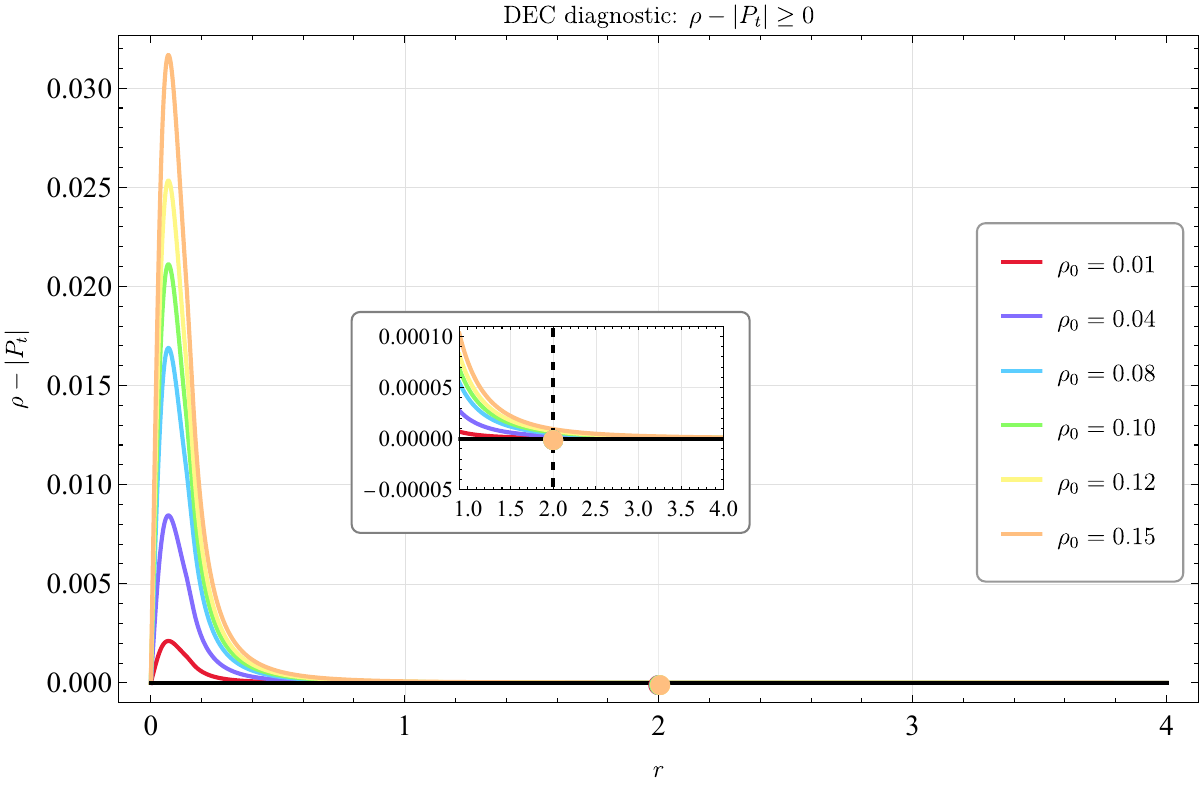}
\includegraphics[scale=0.4]{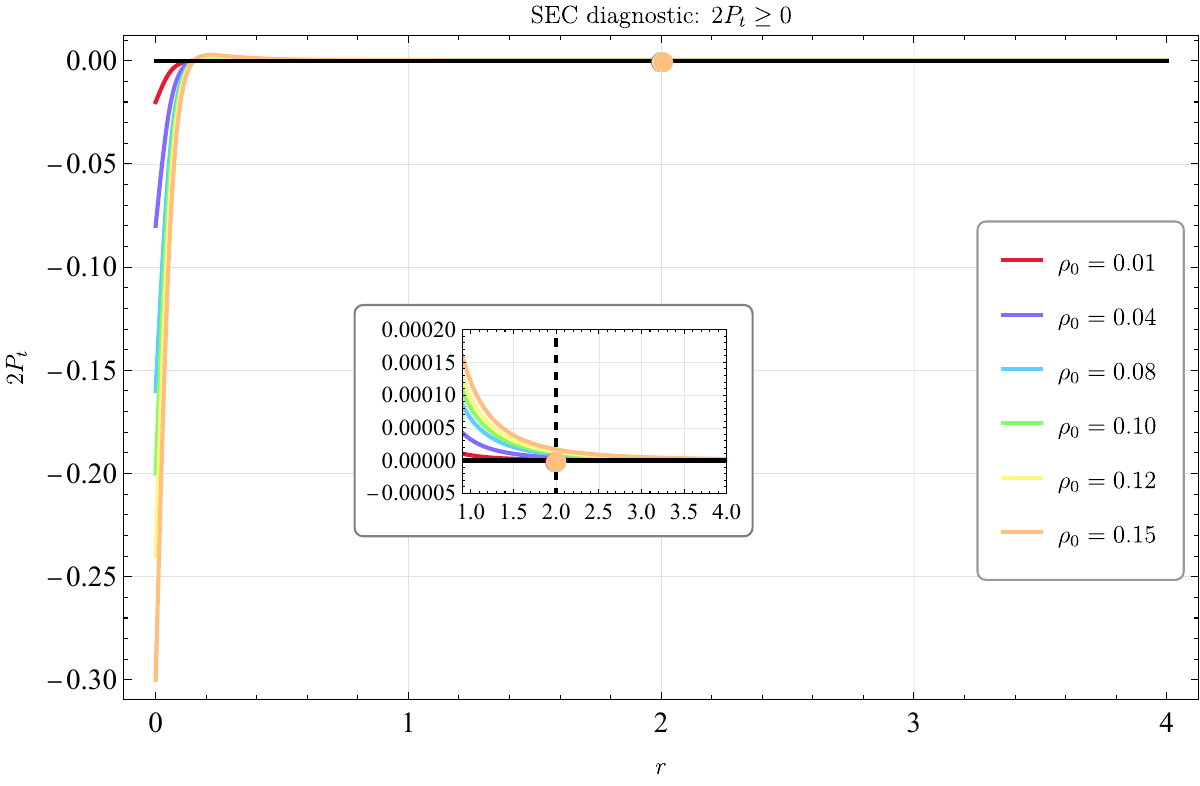}
\end{center}
\setlength{\abovecaptionskip}{-0.1cm}
\setlength{\belowcaptionskip}{1.8cm}
\caption{Energy-condition diagnostics for the Burkert model in a scan over $\rho_0$.
We fix $M=1$ and $r_0=0.1$, and vary $\rho_0\in\{0.01,0.04,0.08,0.10,0.12,0.15\}$.
The plotted quantities are $\rho+P_t$ (WEC), $\rho-|P_t|$ (DEC), and $2P_t=\rho+P_r+2P_t$ (SEC).
The horizontal line denotes the zero baseline separating the allowed ($\ge 0$) and disallowed ($<0$) regions.
Colored dots mark the horizon locations $r_h$ obtained from $\mathcal{F}(r_h)=0$.
The inset highlights the near-horizon behavior and confirms that the exterior region satisfies the required energy conditions.}
\label{fig:ec_scan_rho0}
\end{figure}
For the dominant energy condition, we note that
\begin{equation}
\rho-P_t=\rho_0\,\frac{x^3+2x^2+3x+4}{2(1+x)^2(1+x^2)^2}>0,
\end{equation}
which guarantees $\rho\ge |P_t|$. Together with $\rho-|P_r|=0$, this shows that the dominant energy condition is also satisfied throughout the spacetime region under consideration.

The strong energy condition depends on the sign of
\begin{equation}
\rho+P_r+2P_t = 2P_t ,
\end{equation}
so that
\begin{equation}
\rho+P_r+2P_t \ge 0
\quad\Longleftrightarrow\quad
x^3-x-2\ge 0
\quad\Longleftrightarrow\quad
r\ge r_\star\equiv x_\star r_0,
\end{equation}
where $x_\star\simeq 1.52138$ is the unique positive root of $x^3-x-2=0$. Hence, the SEC may be violated only in the inner region $r<r_\star$ and is restored for sufficiently large radii.

For the parameter range displayed in Figs.~2 and 3, namely $r_0\le 0.15$, one has
\begin{equation}
r_\star=x_\star r_0 \le 0.228,
\end{equation}
whereas the event horizon is located near $r_h\simeq 2$. Therefore, any possible violation of the strong energy condition is hidden well inside the event horizon. In particular, all three energy conditions are satisfied in the physically relevant exterior domain $r\ge r_h$ used in our quasinormal-mode and time-domain analyses.

This result is important for the interpretation of our perturbative study: although the effective anisotropic fluid may exhibit a mild SEC violation deep in the interior, the effective source remains physically well behaved outside the horizon, where the ringdown signal is determined. The behavior of the energy-condition diagnostics is shown in Figs.~\ref{fig:ec_scan_r0} and \ref{fig:ec_scan_rho0}.

\section{Axial gravitational perturbation}

We next study axial gravitational perturbations on the Burkert black hole background. To this end, we introduce a small perturbation $h_{\mu\nu}$ around the background metric $\bar g_{\mu\nu}$,
\begin{equation}
g_{\mu\nu}=\bar g_{\mu\nu}+h_{\mu\nu},
\qquad |h_{\mu\nu}|\ll |\bar g_{\mu\nu}|.
\end{equation}
The corresponding Christoffel symbols can be expanded as
\begin{equation}
\Gamma^{\lambda}_{\mu\nu}
=
\bar{\Gamma}^{\lambda}_{\mu\nu}
+
\delta \Gamma^{\lambda}_{\mu\nu},
\end{equation}
where
\begin{equation}
\delta \Gamma^{\lambda}_{\mu\nu}
=
\frac12 \bar g^{\lambda\beta}
\left(
h_{\mu\beta;\nu}
+
h_{\nu\beta;\mu}
-
h_{\mu\nu;\beta}
\right),
\end{equation}
and the symbol ``$;$'' denotes the covariant derivative with respect to the background metric $\bar g_{\mu\nu}$.

The Ricci tensor is then decomposed as
\begin{equation}
R_{\mu\nu}
=
\bar R_{\mu\nu}
+
\delta R_{\mu\nu},
\end{equation}
with
\begin{equation}
\delta R_{\mu\nu}
=
\delta \Gamma^{\lambda}_{\mu\nu;\lambda}
-
\delta \Gamma^{\lambda}_{\mu\lambda;\nu}.
\label{deltaRicci}
\end{equation}

In the present treatment we adopt a frozen-effective-source approximation:
the independent perturbations of the effective matter variables are neglected,
so that the axial dynamics is described by the metric perturbations on the
fixed anisotropic background. This approximation is best justified when the
effective matter contribution in the strong-field region is small compared
with the central black-hole mass, namely when
\begin{equation}
\eta(r_c)=\frac{m_h(r_c)}{M}\ll 1 .
\end{equation}
The larger-$r_0$ cases considered below should therefore be interpreted as a
phenomenological extension to stronger effective environments rather than as
a controlled weak-environment expansion.

Because the background is not vacuum, the axial perturbation equations should
not be identified with the vacuum condition $\delta R_{\mu\nu}=0$ without
qualification. Instead, after using the background Einstein equations and the
frozen-source closure, the odd-parity metric perturbations can be combined
into a generalized Regge--Wheeler equation. The matter contribution then
appears explicitly in the effective potential through the radial dependence
of the mass function.

For odd-parity perturbations, we adopt the Regge--Wheeler gauge and write
\begin{equation}
h_{\mu\nu}^{\mathrm{odd}}
=
\sum_{l,m}
\begin{pmatrix}
0 & 0 &
-\dfrac{h_{0}(t,r)}{\sin\theta}\partial_{\phi}Y_{lm}
&
h_{0}(t,r)\sin\theta\,\partial_{\theta}Y_{lm}
\\[0.3cm]
0 & 0 &
-\dfrac{h_{1}(t,r)}{\sin\theta}\partial_{\phi}Y_{lm}
&
h_{1}(t,r)\sin\theta\,\partial_{\theta}Y_{lm}
\\[0.3cm]
-\dfrac{h_{0}(t,r)}{\sin\theta}\partial_{\phi}Y_{lm}
&
-\dfrac{h_{1}(t,r)}{\sin\theta}\partial_{\phi}Y_{lm}
&
0 & 0
\\[0.3cm]
h_{0}(t,r)\sin\theta\,\partial_{\theta}Y_{lm}
&
h_{1}(t,r)\sin\theta\,\partial_{\theta}Y_{lm}
&
0 & 0
\end{pmatrix},
\end{equation}
where $Y_{lm}(\theta,\phi)$ denotes the spherical harmonic.

To simplify the subsequent derivation, it is convenient to introduce the function
\begin{equation}
m(r)=M+\pi\rho_0 r_0^3
\left[
2\ln\left(1+\frac{r}{r_0}\right)
+\ln\left(1+\frac{r^2}{r_0^2}\right)
-2\arctan\left(\frac{r}{r_0}\right)
\right],
\end{equation}
so that the metric function can be written in the compact form
\begin{equation}
\mathcal{F}(r)=1-\frac{2m(r)}{r}.
\end{equation}
This representation is particularly useful because it makes the spacetime formally resemble the standard Schwarzschild-like form, while all effects of the surrounding matter distribution are encoded in the radial dependence of \(m(r)\). In this way, the geometric contribution from the Burkert-density effective source can be incorporated into the perturbation analysis in a transparent manner. Under the source-free gravitational perturbation assumption, the nontrivial axial perturbation equations are then obtained from the \((t\phi)\), \((r\phi)\), and \((\theta\phi)\).

From $\delta R_{t\phi}=0$, we obtain
\begin{equation}
\frac{1}{\mathcal{F}(r)}\frac{\partial h_0}{\partial t}
-\mathcal{F}(r)\frac{\partial h_1}{\partial r}
-\frac{2\left[m(r)-r\,m'(r)\right]}{r^2}h_1=0.
\label{Rtphi}
\end{equation}

From $\delta R_{r\phi}=0$, one finds
\begin{equation}
\frac{1}{\mathcal F(r)}
\left(
\frac{\partial^2 h_1}{\partial t^2}
-
\frac{\partial^2 h_0}{\partial t\,\partial r}
\right)
+
\frac{2}{r\mathcal F(r)}\frac{\partial h_0}{\partial t}
+
\frac{(l-1)(l+2)}{r^2}h_1
=0 .
\label{Rrphi}
\end{equation}

From $\delta R_{\theta\phi}=0$, we obtain
\begin{equation}
\left[
\frac{1}{\mathcal F(r)}\frac{\partial h_0}{\partial t}
-\frac{\partial}{\partial r}\!\left(\mathcal F(r)h_1\right)
\right]
\left[
\cos\theta\,\partial_{\theta}Y_{lm}
-\sin\theta\,\partial_{\theta}^{2}Y_{lm}
+\frac{1}{\sin\theta}\partial_{\phi}^{2}Y_{lm}
\right]
=0 .
\label{Rthetaphi}
\end{equation}

Next, we introduce the tortoise coordinate
\begin{equation}
\frac{dr_*}{dr}=\frac{1}{\mathcal F(r)},
\end{equation}
and define the master variable
\begin{equation}
\Psi(t,r)=\frac{\mathcal F(r)}{r}h_1(t,r).
\end{equation}
With these definitions, Eqs.~\eqref{Rtphi}--\eqref{Rthetaphi} combine into a Regge--Wheeler-type wave equation
\begin{equation}\label{eq42}
-\frac{\partial^2\Psi}{\partial t^2}
+
\frac{\partial^2\Psi}{\partial r_*^2}
-
V_{\mathrm{axial}}(r)\Psi
=0 ,
\end{equation}
where the effective potential takes the explicit form
\begin{equation}
V_{\mathrm{axial}}(r)
=
\mathcal F(r)
\left[
\frac{\ell(\ell+1)}{r^2}
-\frac{6M}{r^3}
-\frac{6\pi \rho_{0}r_{0}^{3}}{r^3}
\left(
2\ln\left(1+\frac{r}{r_{0}}\right)
+\ln\left(1+\frac{r^{2}}{r_{0}^{2}}\right)
-2\arctan\left(\frac{r}{r_{0}}\right)
\right)
+\frac{8\pi \rho_{0}r_{0}^{3}}{(r+r_{0})(r^{2}+r_{0}^{2})}
\right].
\label{Vexplicit}
\end{equation}

Although the mass function grows logarithmically at very large radius,
the metric function satisfies
\[\lim_{r\to\infty}\mathcal{F}(r)=1,\]
indicating that the spacetime is asymptotically flat.
Consequently,
\[
V_{\rm axial}(r)\rightarrow 0,
\qquad r\rightarrow\infty.
\]
In this limit, Eq.~(\ref{eq42}) reduces to the free-wave equation
\[
-\frac{\partial^2\Psi}{\partial t^2}
+
\frac{\partial^2\Psi}{\partial r_*^2}
=0 ,
\]
which admits the standard ingoing and outgoing plane-wave solutions.
Therefore, despite the logarithmic growth of the mass function,
the usual QNM boundary conditions remain well defined: purely ingoing waves at the horizon and purely outgoing waves at spatial infinity.

\section{The methods}
To analyze the perturbation dynamics from both frequency-domain and time-domain perspectives, we employ three complementary numerical approaches in this work. In the frequency domain, the QNM frequencies are computed by using the Chebyshev pseudospectral method and the Borel--Pad\'e method. In the time domain, the perturbation evolution is obtained through a characteristic integration scheme in double-null coordinates. The resulting waveform is then fitted in the ringdown regime by the Prony method, which allows us to extract the dominant complex frequencies directly from the time-domain data. In this way, the results obtained from the frequency-domain and time-domain analyses can be compared with each other and used for mutual validation.
\subsection{Pseudospectral method}
\label{subsec:pseudo}
The pseudospectral method \cite{Konoplya:2024lch,Jaramillo:2020tuu,Yang:2025hqk,Jansen:2017oag}  transforms the original boundary-value problem into a finite-dimensional eigenvalue problem on a compact domain. After compactifying the semi-infinite radial direction to a finite interval, one discretizes the radial equation on a Chebyshev--Lobatto collocation grid and represents derivatives by differentiation matrices, so that the QNM frequencies can be extracted from the resulting generalized matrix eigenvalue problem.
As a first step in the pseudospectral discretization, we introduce the compactified radial coordinate
\begin{equation}
    \sigma=\frac{r_h}{r}, \qquad \sigma\in[0,1],
\end{equation}
so that the semi-infinite radial domain is mapped onto the finite interval $[0,1]$, with the event horizon located at $\sigma=1$ and spatial infinity at $\sigma=0$. In terms of the compactified coordinate $\sigma$, the axial gravitational perturbation equation takes the form
\begin{equation}
w(\sigma)\,\partial_{\tau\tau}\Psi
-2\gamma(\sigma)\,\partial_{\tau\sigma}\Psi
-p(\sigma)\,\partial_{\sigma\sigma}\Psi
-\gamma'(\sigma)\,\partial_\tau\Psi
-p'(\sigma)\,\partial_\sigma\Psi
+q(\sigma)\Psi=0,
\label{eq:pseudo-master}
\end{equation}
where the coefficient functions are
\begin{equation}
p(\sigma)=\sigma^2 F(\sigma),
\qquad
\gamma(\sigma)=-1+2F(\sigma)S(\sigma),
\qquad
w(\sigma)=-\frac{4S(\sigma)\bigl[F(\sigma)S(\sigma)-1\bigr]}{\sigma^2},
\end{equation}
and
\begin{equation}
q(\sigma)=\ell(\ell+1)+2\bigl[F(\sigma)-1\bigr]+\sigma F'(\sigma).
\label{eq:qaxial-short}
\end{equation}
Here $F(\sigma)\equiv F(r_h/\sigma)$, and the prime denotes the derivative with respect to \(\sigma\).

Because the asymptotic expansion of the background near null infinity ($\sigma=0$) contains logarithmic terms, we introduce a regularizing factor $S(\sigma)$ such that the coefficient functions are finite at $\sigma=0$. In particular, if
\begin{equation}
F(\sigma)=1+a_\infty \sigma \ln \sigma+b_\infty \sigma+e_\infty \sigma^2+\cdots,
\end{equation}
we choose
\begin{equation}
S(\sigma)=
1-a_\infty\sigma\ln\sigma-b_\infty\sigma
+a_\infty^2\sigma^2\ln^2\sigma
+2a_\infty b_\infty\sigma^2\ln\sigma
-e_\infty\sigma^2,
\end{equation}
which regularizes the equation at $\sigma=0$ and reduces to the standard Schwarzschild form in the vacuum limit.

With the harmonic ansatz
\begin{equation}
\Psi(\tau,\sigma)=e^{-i\omega\tau}\phi(\sigma),
\end{equation}
Eq.~\eqref{eq:pseudo-master} becomes a quadratic eigenvalue problem,
\begin{equation}
\bigl(L_1-i\omega L_2+\omega^2 I\bigr)\bm{\phi}=0,
\label{eq:quadratic-eigen-short}
\end{equation}
where
\begin{equation}
L_1=\mathrm{diag}(\mathcal{C})D^{(2)}
+\mathrm{diag}(\mathcal{E})D^{(1)}
+\mathrm{diag}(\mathcal{W}),
\qquad
L_2=\mathrm{diag}(\mathcal{A})D^{(1)}
+\mathrm{diag}(\mathcal{B}).
\end{equation}
Here \(D^{(1)}\) and \(D^{(2)}\) are the first- and second-order Chebyshev--Lobatto differentiation matrices on the collocation grid
\begin{equation}
\sigma_j=\frac{1-\cos(j\pi/N)}{2},\qquad j=0,1,\dots,N,
\end{equation}
and \(\mathrm{diag}(\mathcal C)\), \(\mathrm{diag}(\mathcal E)\), \(\mathrm{diag}(\mathcal W)\), \(\mathrm{diag}(\mathcal A)\), and \(\mathrm{diag}(\mathcal B)\) denote diagonal matrices whose entries are the corresponding coefficient functions evaluated at the collocation points. Thus, for example,
\begin{equation}
\mathrm{diag}(\mathcal C)=
\mathrm{diag}\bigl(\mathcal C(\sigma_0),\mathcal C(\sigma_1),\dots,\mathcal C(\sigma_N)\bigr),
\end{equation}
so that the coefficient functions enter the discretized equations as pointwise multiplication operators on the collocation grid.

At the two endpoints, we impose the coefficient functions analytically rather than by taking direct symbolic limits. At $\sigma=0$, one has
\begin{equation}
\mathcal{C}(0)=0,\qquad
\mathcal{E}(0)=0,\qquad
\mathcal{W}(0)=-\frac{\ell(\ell+1)}{4b_\infty^2},\qquad
\mathcal{A}(0)=\frac{1}{2b_\infty^2},\qquad
\mathcal{B}(0)=0,
\end{equation}
whereas at $\sigma=1$, writing $s_h=S(1)$ and $f_h=F'(1)$, we use
\begin{equation}
\mathcal{C}(1)=0,\qquad
\mathcal{E}(1)=\frac{f_h}{4s_h},\qquad
\mathcal{W}(1)=-\frac{\ell(\ell+1)-2+f_h}{4s_h},
\end{equation}
\begin{equation}
\mathcal{A}(1)=-\frac{1}{2s_h},\qquad
\mathcal{B}(1)=\frac{f_h}{2}.
\end{equation}

The discretization described above reduces the perturbation equation to a finite-dimensional matrix eigenvalue problem, whose complex eigenvalues give the quasinormal frequencies. In practice, we solve this problem at several spectral resolutions and identify the physically relevant frequencies from those eigenvalues that remain stable as the number of collocation points is increased. 

\subsection{Numerical methods for time-domain profiles}

For the time-domain analysis, we use the numerical characteristic integration scheme developed by Gundlach, Price, and collaborators \cite{price,price2}. In light-cone coordinates,
\begin{equation}
\begin{array}{l}
u=t-r_*, \\
v=t+r_*,
\end{array}
\end{equation}
the wave equation for each perturbation sector can be written as
\begin{equation}\label{eq15}
\frac{\partial^{2}}{\partial u \partial v} \psi(u, v)+\frac{1}{4}V(r_*) \psi(u, v)=0.
\end{equation}
We discretize this equation according to
\begin{equation}
\psi_{N}=\psi_{E}+\psi_{W}-\psi_{S}-\Delta u \Delta v V(r_*)\left(\frac{\psi_{W}+\psi_{E}}{8}\right)+\mathcal{O}\left(\Delta^{4}\right).
\end{equation}
where the grid points are defined by
\begin{equation}
S=(u,v),\qquad W=(u+\Delta u,v),\qquad E=(u,v+\Delta v),\qquad N=(u+\Delta u,v+\Delta v).
\end{equation}

As initial data, we choose a Gaussian pulse \cite{Moderski:2001gt},
\begin{equation}
\begin{array}{l}
\psi\left(u=u_{0}, v\right)=\exp \left[-\frac{\left(v-v_{c}\right)^{2}}{2 \sigma^{2}}\right], \\
\psi\left(u, v=v_{0}\right)=0.
\end{array}
\end{equation}
Each new value $\psi_N$ is then obtained from the three neighboring points already known on the grid. Repeating this procedure yields the full time evolution and allows us to extract the ringdown profiles for the Burkert black hole.

\section{The QNM frequencies and ringdown of the BH surrounded by the Burkert profile}\label{rbh_pi}

We now present the fundamental quasinormal frequencies of the Burkert black hole and compare the values obtained from the pseudospectral method \cite{Konoplya:2024lch,Jaramillo:2020tuu,Yang:2025hqk,Jansen:2017oag}, Borel summation method \cite{Hatsuda:2019eoj,Sulejmanpasic:2016fwr}, and Prony method \cite{Konoplya:2011qq} extraction from time-domain profiles, as shown in TABLE~\ref{qnf-rho0} and TABLE~\ref{qnf-r0}. The excellent agreement among these methods provides a strong consistency check and confirms the reliability of both the frequency domain and time-domain analyzes.

From TABLE~\ref{qnf-rho0}, we observe that as the density $\rho_0$ increases, both the oscillation frequency $\mathrm{Re}(\omega)$ and the damping rate $|\mathrm{Im}(\omega)|$ decrease monotonically. This implies that a larger effective Burkert-density contribution leads to slower oscillations and longer-lived ringdown signals. It is also worth emphasizing that the dependence on $r_0$ is significantly stronger than that on $\rho_0$, as shown in TABLE~\ref{qnf-r0}. This indicates that the size of the constant-density core of the Burkert-density effective source plays a dominant role in shaping the dynamical response of the spacetime. This can be understood from the fact that, in the Burkert profile, $r_0$ represents the core radius of the Burkert density profile, which determines the size of the central constant-density region. Increasing $r_0$ effectively extends this core region and shifts the transition between the inner flat profile and the outer decaying profile to larger radii.
\begin{table}[b]
\centering
\caption{Fundamental quasinormal mode frequencies ($n=0$) for different values of $\rho_0$. We fix \(M=1\), \(r_0=0.1\), \(\ell=2\). }
\renewcommand{\arraystretch}{1.6}
\setlength{\tabcolsep}{18pt}
\begin{tabular}{cccc}
\hline\hline
$\rho_0$ & $\omega$ (Pseudospectral) & $\omega$ (Borel--Pad\'e) & $\omega$ (Prony) \\
\hline\hline
0.1 & $0.3724420 - 0.0886172\,i$ & $0.3724420 - 0.0886171\,i$ & $0.3726906 - 0.0885938\,i$ \\
0.2 & $0.3712176 - 0.0882739\,i$ & $0.3712176 - 0.0882738\,i$ & $0.3714645 - 0.0882503\,i$ \\
0.3 & $0.3699985 - 0.0879325\,i$ & $0.3699984 - 0.0879324\,i$ & $0.3702448 - 0.0879060\,i$ \\
0.4 & $0.3687846 - 0.0875929\,i$ & $0.3687845 - 0.0875927\,i$ & $0.3690262 - 0.0875716\,i$ \\
0.5 & $0.3675760 - 0.0872552\,i$ & $0.3675758 - 0.0872549\,i$ & $0.3678183 - 0.0872308\,i$ \\
0.6 & $0.3663725 - 0.0869193\,i$ & $0.3663723 - 0.0869190\,i$ & $0.3666104 - 0.0868970\,i$ \\
0.7 & $0.3651742 - 0.0865852\,i$ & $0.3651740 - 0.0865848\,i$ & $0.3654101 - 0.0865629\,i$ \\
0.8 & $0.3639811 - 0.0862529\,i$ & $0.3639809 - 0.0862524\,i$ & $0.3642149 - 0.0862305\,i$ \\
0.9 & $0.3627931 - 0.0859224\,i$ & $0.3627928 - 0.0859219\,i$ & $0.3630248 - 0.0858998\,i$ \\
1.0 & $0.3616102 - 0.0855936\,i$ & $0.3616099 - 0.0855931\,i$ & $0.3618392 - 0.0855707\,i$ \\
\hline\hline
\end{tabular}
\label{qnf-rho0}
\vspace{25pt}
\end{table}

\begin{table}[htbp]
\centering
\caption{Fundamental quasinormal mode frequencies ($n=0$) for different values of $r_0$. We fix \(M=1\), \(\rho_0=0.1\), \(\ell=2\).}
\renewcommand{\arraystretch}{1.6}
\setlength{\tabcolsep}{18pt}
\begin{tabular}{cccc}
\hline\hline
$r_0$ & $\omega$ (Pseudospectral) & $\omega$ (Borel--Pad\'e) & $\omega$ (Prony) \\
\hline\hline
0.1 & $0.3724420 - 0.0886172\,i$ & $0.3724420 - 0.0886171\,i$ & $0.3726906 - 0.0885938\,i$ \\
0.2 & $0.3663682 - 0.0868354\,i$ & $0.3663680 - 0.0868350\,i$ & $0.3666077 - 0.0868123\,i$ \\
0.3 & $0.3539789 - 0.0830941\,i$ & $0.3539781 - 0.0830924\,i$ & $0.3541943 - 0.0830652\,i$ \\
0.4 & $0.3351861 - 0.0773505\,i$ & $0.3351841 - 0.0773452\,i$ & $0.3353793 - 0.0773352\,i$ \\
0.5 & $0.3105627 - 0.0698616\,i$ & $0.3105587 - 0.0698486\,i$ & $0.3106512 - 0.0698564\,i$ \\
0.6 & $0.2810041 - 0.0610683\,i$ & $0.2809979 - 0.0610415\,i$ & $0.2810721 - 0.0610084\,i$ \\
0.7 & $0.2476047 - 0.0515240\,i$ & $0.2475987 - 0.0514788\,i$ & $0.2476305 - 0.0514150\,i$ \\
0.8 & $0.2116959 - 0.0418591\,i$ & $0.2116969 - 0.0417986\,i$ & $0.2116130 - 0.0417998\,i$ \\
0.9 & $0.1749600 - 0.0327285\,i$ & $0.1750039 - 0.0326914\,i$ & $0.1749127 - 0.0326763\,i$ \\
1.0 & $0.1396533 - 0.0248257\,i$ & $0.1396986 - 0.0248075\,i$ & $0.1390611 - 0.0248126\,i$ \\
\hline\hline
\end{tabular}
\label{qnf-r0}
\vspace{25pt}
\end{table}

We emphasize that the larger values of $r_0$ in Table~\ref{qnf-r0}
correspond to strong effective-environment configurations. They are included
to illustrate the qualitative parameter dependence of the spectrum, but the
frozen-effective-source approximation is most reliable for the small-$r_0$
part of the scan, where $\eta(r_c)\ll 1$.

The above behaviors admit a clear geometric interpretation in the eikonal regime. In particular, in the eikonal limit ($l \gg n$), the QNM spectrum is closely related to the properties of the unstable photon sphere via the correspondence \cite{Cardoso:2008bp}
\begin{equation}\label{wln}
\omega_{ln} \approx l\,\Omega_c - i\left(n+\frac{1}{2}\right)|\lambda|,
\end{equation}
where $\Omega_c$ is the angular frequency of the null circular orbit and $\lambda$ is the associated Lyapunov exponent characterizing its instability. Their specific expression can be written as
\begin{equation}
\Omega_c=\frac{\sqrt{F(r_c)}}{r_c}=\frac{\sqrt{F_c}}{r_c},
\label{eq:Omegac}
\end{equation}
and
\begin{equation}
\lambda=\sqrt{\frac{F(r_c)}{2r_c^2}\left[2F(r_c)-r_c^2F''(r_c)\right]}
=\frac{\sqrt{F_c}}{r_c}\sqrt{\Delta_c}.
\label{eq:lambdac}
\end{equation}
Substituting the above two equations into equation (\ref{wln}), we obtain
\begin{equation}
\omega_{ln}\simeq \frac{\sqrt{F_c}}{r_c}
\left[
l-i\left(n+\frac12\right)\sqrt{\Delta_c}
\right],
\end{equation}
where
\begin{equation}
F_c=1-\frac{2M}{r_c}-\frac{2\pi\rho_0 r_0^3}{r_c}H_c,
\end{equation}
\begin{equation}
H_c\equiv
2\ln\!\left(1+\frac{r_c}{r_0}\right)
+\ln\!\left(1+\frac{r_c^2}{r_0^2}\right)
-2\arctan\!\left(\frac{r_c}{r_0}\right),
\end{equation}
and
\begin{equation}
\Delta_c\equiv
1-\frac{4\pi\rho_0 r_0^3 r_c^3(3r_c^2+2r_c r_0+r_0^2)}
{(r_c+r_0)^2(r_c^2+r_0^2)^2}.
\end{equation}
The photon sphere radius $r_c$ is determined by
\begin{equation}
r_c-3M-\pi\rho_0 r_0^3
\left(
3H_c-\frac{4r_c^3}{(r_c+r_0)(r_c^2+r_0^2)}
\right)=0.
\end{equation}
In this limit, the QNM spectrum is governed by the properties of the unstable photon sphere: the real part is set by the angular velocity $\Omega_c$ of the null circular orbit, whereas the imaginary part is determined by its instability timescale encoded in the Lyapunov exponent $\lambda$.

\begin{figure}[b]
\vspace{25pt}
\centering
\makebox[\textwidth][c]{
\includegraphics[scale=0.55]{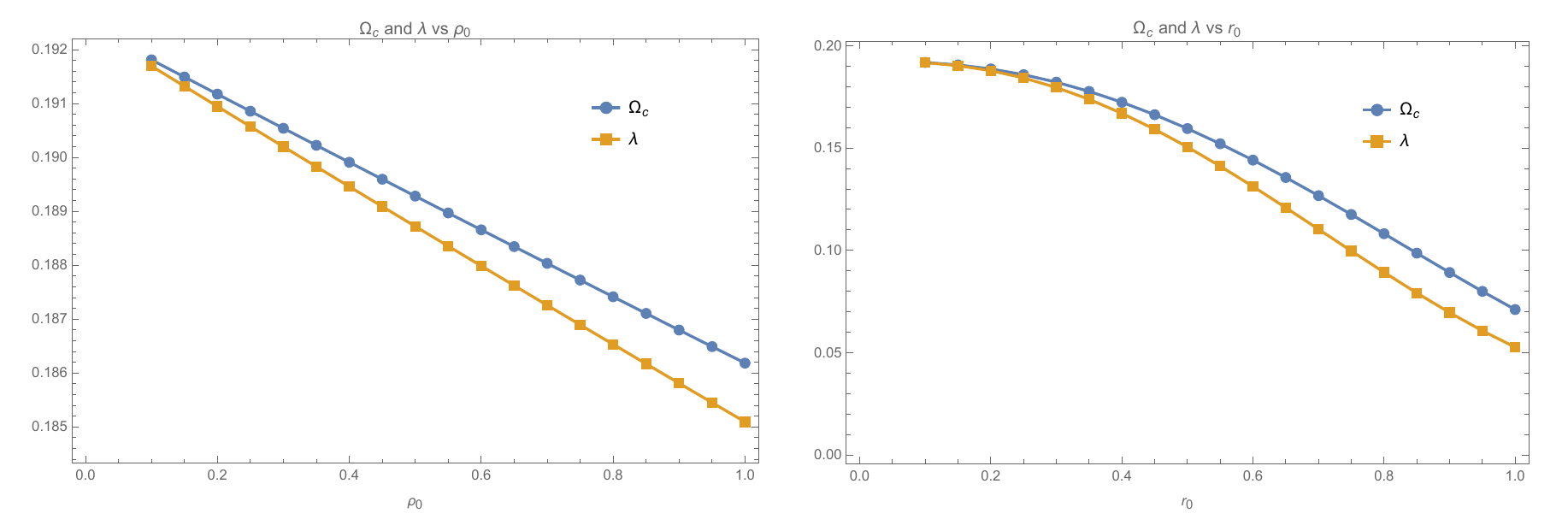}}
\vspace{-20pt}
\caption{Angular frequency and Lyapunov exponent of the photon sphere as functions of the Burkert-density effective-source parameters. Left: $\Omega_c$ and $\lambda$ versus $\rho_0$ for fixed \(M=1\), \(r_0=0.1\). Right: $\Omega_c$ and $\lambda$ versus $r_0$ for fixed \(M=1\), \(\rho_0=0.1\).}
\label{omage-lambda}
\end{figure}

Fig.~\ref{omage-lambda} shows the behavior of the photon-sphere angular frequency $\Omega_c$ and the Lyapunov exponent $\lambda$ as functions of the effective-source parameters $\rho_0$ and $r_0$. Both quantities decrease monotonically with increasing $\rho_0$ and $r_0$, while the dependence on $r_0$ is significantly stronger, indicating that the core size of the Burkert halo has a more pronounced impact on the photon-sphere properties than the density.
From this perspective, the role of the effective Burkert-density source is straightforward. The Burkert-density effective matter distribution modifies the background geometry and thereby changes both $\Omega_c$ and $\lambda$. As $\rho_0$ or $r_0$ increases, the angular velocity $\Omega_c$ decreases, leading to a smaller oscillation frequency, while the Lyapunov exponent $\lambda$ is also reduced, implying weaker instability of the null circular orbit and hence a smaller damping rate. This naturally explains the systematic decrease of both $\mathrm{Re}(\omega)$ and $|\mathrm{Im}(\omega)|$ observed in TABLE~\ref{qnf-rho0} and TABLE~\ref{qnf-r0}.

It is also worth emphasizing that the dependence on $r_0$ is significantly stronger than that on $\rho_0$. This suggests that the global structure of the effective Burkert-density distribution, rather than its local density alone, plays a more crucial role in determining the QNM spectrum. Since $r_0$ controls the spatial extent over which dark matter modifies the geometry, its impact propagates more efficiently to the photon sphere region, thereby inducing larger deviations in the ringdown frequencies.

From an observational perspective, these results indicate that dark-matter-inspired effective environments can leave characteristic imprints on gravitational wave ringdown signals. In particular, the simultaneous decrease in oscillation frequency and damping rate may serve as a potential signature of environmental effects surrounding black holes. Therefore, precise measurements of QNMs could, in principle, provide a probe of effective environmental corrections in strong-gravity regimes. A direct interpretation in terms of microscopic particle dark matter would require a more complete relativistic halo model, including an inner edge, an outer cutoff, and matter perturbations.

\begin{figure}[htbp]
\vspace{25pt}
\centering
\makebox[\textwidth][c]{
\includegraphics[scale=0.65]{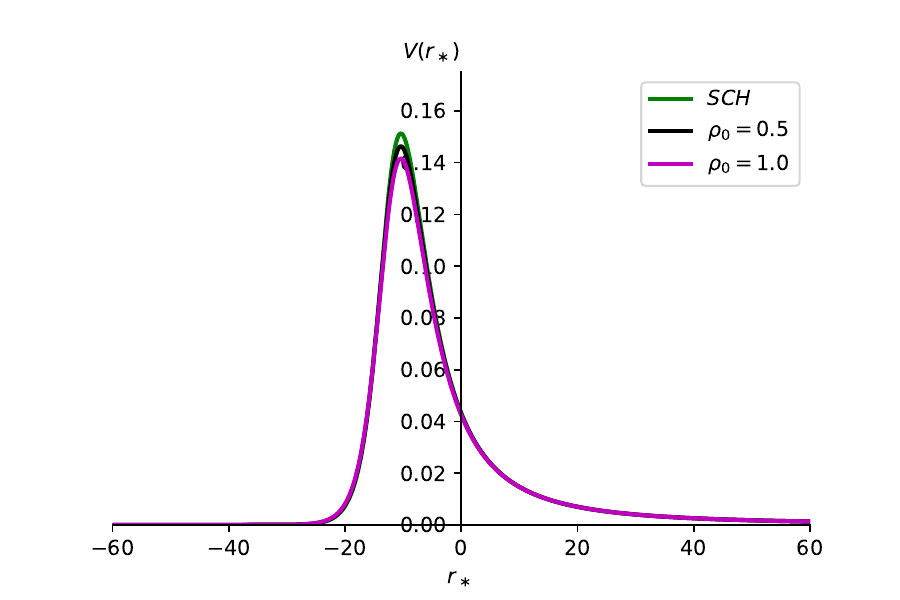}
\hspace{-1.1cm}
\includegraphics[scale=0.65]{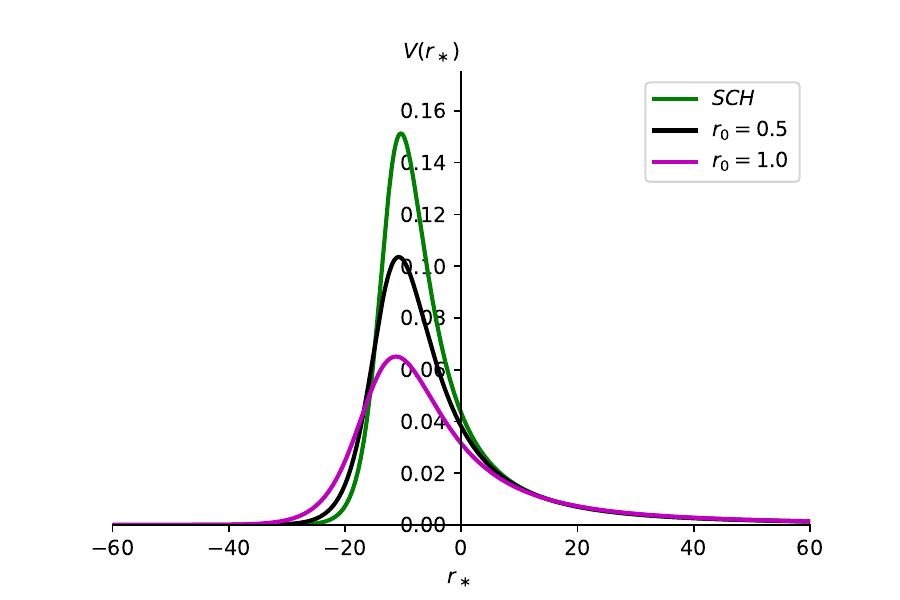}}
\vspace{-20pt}
\caption{Effective potential $V(r_*)$ for gravitational perturbations in the Burkert-density effective background. Left panel: $V(r_*)$ for different values of $\rho_0$ with $M=1$, $r_{0}=0.1$, $\ell=2$. Right panel: $V(r_*)$ for different values of $r_0$ with fixed $M=1$, $\rho_{0}=0.1$, $\ell=2$. In both panels, the Schwarzschild case (SCH) is shown for comparison.}
\label{V}
\vspace{20pt}
\end{figure}

\begin{figure}[htbp]
\centering
\makebox[\textwidth][c]{
\includegraphics[scale=0.65]{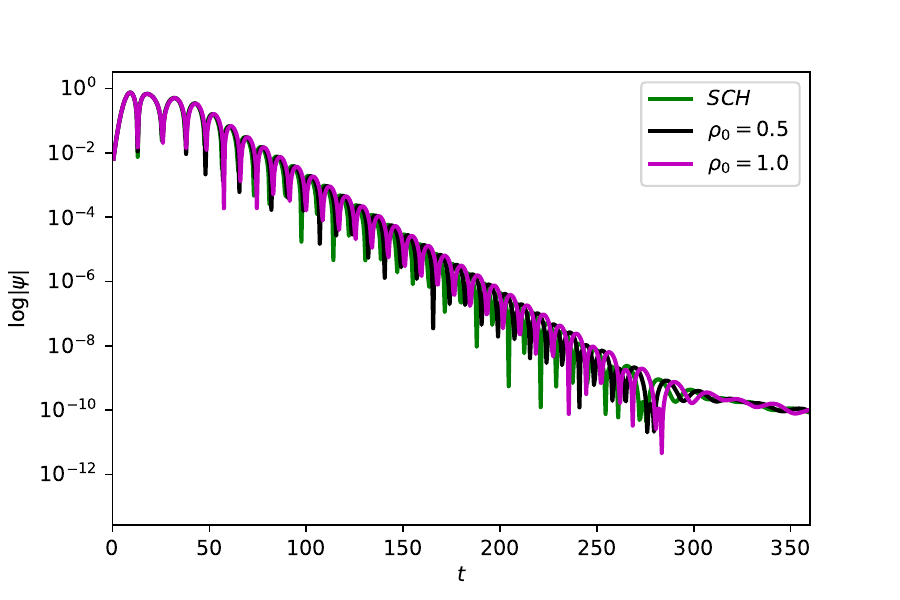}
\hspace{-1.1cm}
\includegraphics[scale=0.65]{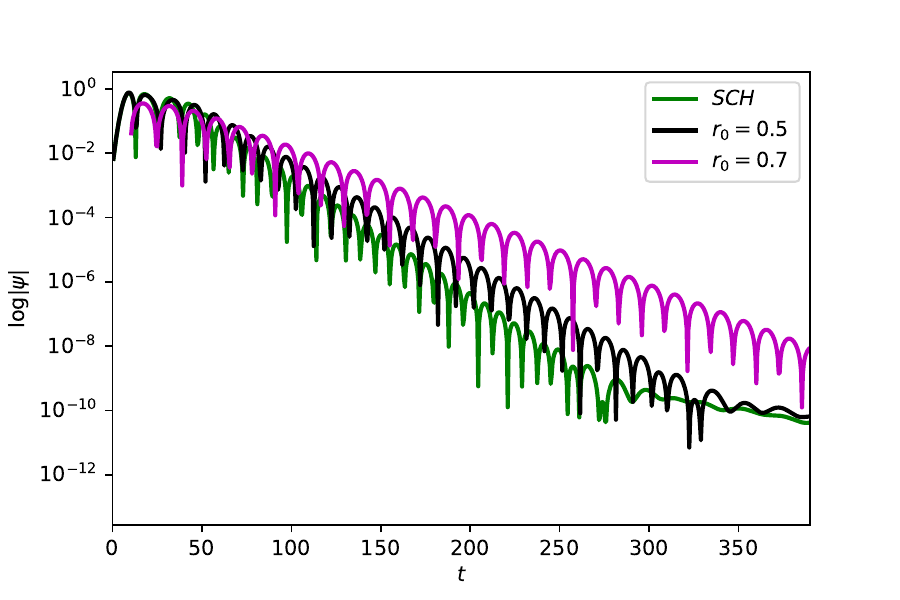}}
\vspace{-20pt}
\caption{Time-domain profiles of the gravitational perturbation for different values of $\rho_{0}$ (left) and $r_{0}$ (right), with $M=1$, $r_{0}=0.1$, $\ell=2$ in the left panel, and $M=1$, $\rho_{0}=0.1$, $\ell=2$ in the right panel, respectively.}
\label{td1}
\end{figure}

\begin{figure}[htbp]
\centering
\makebox[\textwidth][c]{
\includegraphics[scale=0.65]{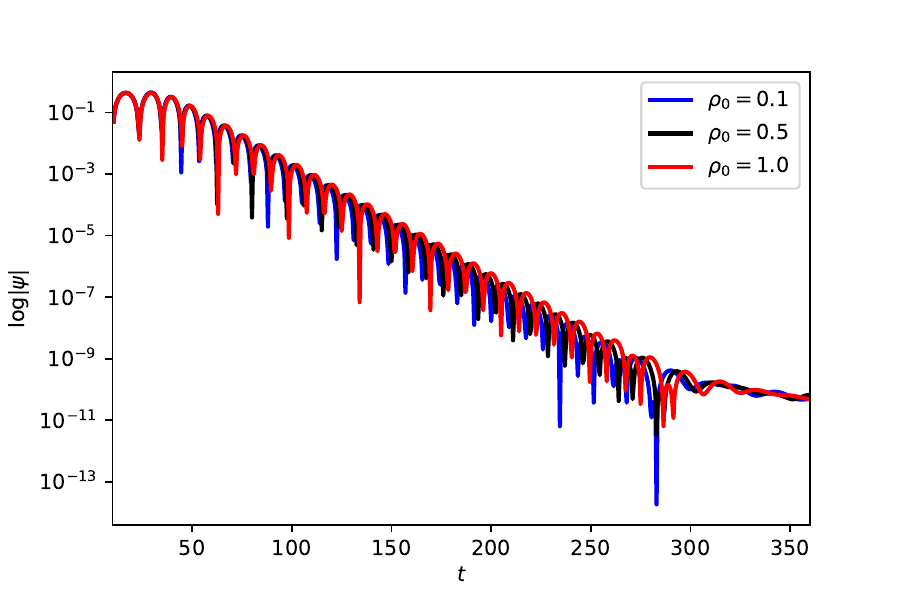}
\hspace{-1.1cm}
\includegraphics[scale=0.65]{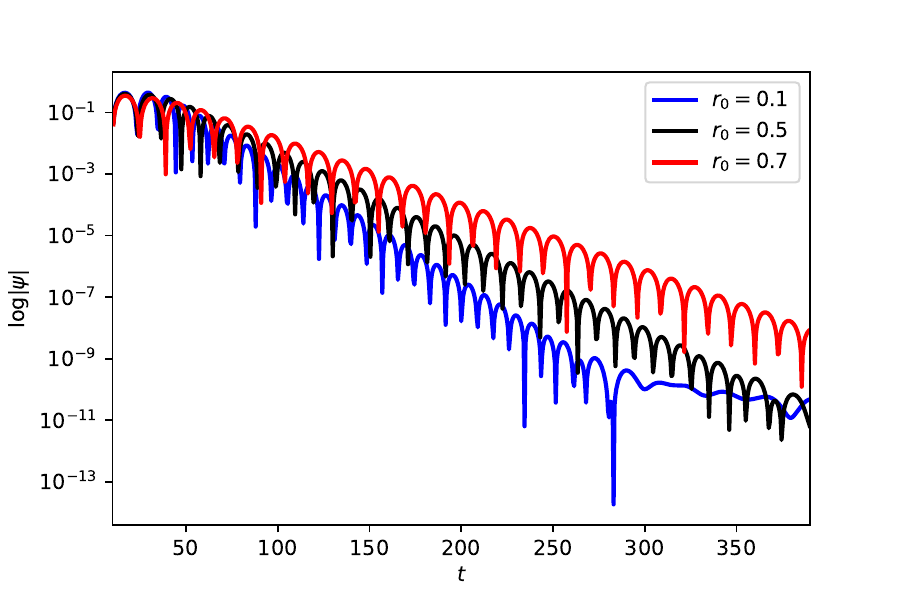}}
\vspace{-20pt}
\caption{Time-domain profiles of the gravitational perturbation for different values of $\rho_{0}$ (left) and $r_{0}$ (right), with $M=1$, $r_{0}=0.1$, $\ell=2$ in the left panel, and $M=1$, $\rho_{0}=0.1$, $\ell=2$ in the right panel, respectively.}
\label{td2}
\vspace{25pt}
\end{figure}

Figs.~\ref{V}--\ref{td2} further illustrate how the Burkert-density effective matter distribution modifies the gravitational perturbation dynamics at both the level of the effective potential and in the time domain.

Fig.~\ref{V} displays the effective potential $V(r_*)$ for gravitational perturbations. In the left panel, varying $\rho_0$ while keeping $r_0$ fixed produces only a mild suppression of the potential barrier relative to the Schwarzschild case. The peak height decreases slightly as $\rho_0$ increases, while the overall single-barrier structure is preserved. By contrast, the right panel shows that the dependence on the core radius $r_0$ is much stronger: increasing $r_0$ significantly lowers the barrier and makes it broader, indicating a more substantial modification of the background geometry in the photon-sphere region. In both panels, the potential remains single-peaked, with no evidence of a secondary barrier or potential well. This implies that the Burkert-density effective background considered here does not generate the kind of trapping structure typically associated with gravitational wave echoes \cite{Dong:2020odp,Huang:2021qwe,DuttaRoy:2019hij}.

The behavior of the effective potential in Fig.~\ref{V} provides a clear qualitative explanation for the QNM trends reported in TABLE~\ref{qnf-rho0} and TABLE~\ref{qnf-r0}. In the left panel, increasing $\rho_0$ only mildly lowers the peak of the barrier relative to the Schwarzschild case, whereas in the right panel increasing $r_0$ produces a much more pronounced suppression of the peak. Since the characteristic scale and shape of the potential barrier are closely related to the oscillation frequency and damping rate of the ringdown modes, a lower barrier naturally tends to produce smaller oscillation frequencies and weaker damping. This is consistent with our frequency domain results. Moreover, the more pronounced change in the barrier structure induced by $r_0$ explains why the QNM spectrum is considerably more sensitive to the core radius than to the density amplitude $\rho_0$.

This interpretation based on the effective potential and frequency domain results is further confirmed in the time domain. Fig.~\ref{td1} compares the time-domain profiles with the Schwarzschild case. In the left panel, increasing $\rho_0$ leads to a slightly longer oscillation period and a slightly slower decay, but the overall deviation from Schwarzschild remains modest. In the right panel, varying $r_0$ produces a far more pronounced effect: the oscillation period increases substantially and the decay becomes much slower as $r_0$ grows. Thus, the time-domain signal directly reflects the same pattern already seen in the effective potential and in the QNM tables, namely that the core radius $r_0$ is the parameter that dominantly controls the ringdown modification.

Fig.~\ref{td2} emphasizes the parameter dependence more clearly by comparing only the Burkert-density effective cases. The left panel shows that, as $\rho_0$ increases from smaller to larger values, the waveform shifts toward lower oscillation frequency and slightly reduced damping. The right panel shows an even stronger monotonic dependence on $r_0$: larger $r_0$ produces markedly slower oscillations and much more persistent late-time ringing. In other words, the ringdown signal becomes longer-lived as the Burkert halo becomes more extended. This behavior is qualitatively consistent with the decrease of the photon sphere angular frequency $\Omega_c$ and the Lyapunov exponent $\lambda$ in the eikonal regime, and therefore supports the geometric interpretation discussed above.

Taken together, Figs.~\ref{V}--\ref{td2} establish a coherent physical picture. The Burkert-density effective halo modifies the geometry near the photon sphere, lowers the effective barrier, reduces the characteristic oscillation scale, and weakens the instability of the relevant null orbit. As a result, the gravitational perturbation exhibits lower-frequency and more slowly damped ringdown modes. The effect of $\rho_0$ is comparatively mild, whereas the effect of $r_0$ is much stronger, showing that the size of the central core plays a more important role than the density amplitude in determining the dynamical response of the spacetime.

\section{Conclusion and outlook}\label{sec:summary}

In this work, we have constructed a static, spherically symmetric black hole spacetime embedded in a Burkert-density effective halo and investigated its ringdown response under axial gravitational perturbations. Starting from the Burkert density profile, we derived the corresponding halo contribution to the background geometry and obtained an analytic Schwarzschild-like metric characterized by the two halo parameters $(\rho_0,r_0)$. The vacuum Schwarzschild solution is smoothly recovered when the halo contribution is switched off, so the model provides a clean and controlled framework in which the influence of a cored dark matter environment can be isolated explicitly.

We then analyzed the perturbative dynamics of this spacetime by studying the effective potential, the quasinormal spectrum, and the time-domain evolution of gravitational perturbations. The results obtained from the pseudospectral method, Borel--Pad\'e treatment, and Prony extraction from time-domain profiles are in very good agreement, providing a strong consistency check on the numerical analysis. Our results show that the presence of the Burkert halo systematically shifts the ringdown spectrum relative to the vacuum case: increasing either the central density $\rho_0$ or the core radius $r_0$ lowers both the oscillation frequency and the damping rate, while the dependence on $r_0$ is significantly stronger. This indicates that the spatial extent of the cored halo plays a more important role than the density amplitude in shaping the dynamical response of the black hole spacetime.

A useful physical interpretation emerges from the eikonal picture. the Burkert-density effective halo modifies the geometry near the photon sphere, reduces both the angular velocity of the unstable null orbit and its Lyapunov exponent, and correspondingly produces lower-frequency and more slowly damped quasinormal ringing. This interpretation is consistent with the behavior of the effective potential, whose barrier becomes lower and broader as the halo parameters increase, especially for larger values of the core radius. The time-domain profiles confirm the same trend, showing that the ringdown becomes progressively longer-lived in a more extended effective environment.

We also examined the physical viability of the effective matter source supporting the geometry. For the parameter range considered in this work, the weak and dominant energy conditions are satisfied in the exterior region relevant to the perturbation problem, while any violation of the strong energy condition is confined inside the event horizon. This result supports the interpretation of the model as a reasonable effective description of a black hole immersed in a cored halo, at least in the domain that governs observable ringdown physics.

From a broader perspective, our analysis provides further evidence that environmental matter distributions can leave nontrivial and, in principle, observable imprints on black hole spectroscopy. Although the present effects are modest for part of the parameter space, they are systematic and physically transparent, making the Burkert-density effective model a useful benchmark for studying how dark-matter-inspired effective environments may affect strong-gravity observables. In this sense, the present work complements ongoing efforts to understand ``dirty'' black holes and to assess how astrophysical environments may enter future precision tests of gravity with ringdown measurements.

We finally stress that the matter sector used in this work should be interpreted as a Burkert-density effective anisotropic source. 
It is inspired by the cored Burkert dark-matter density profile, but it is not a microscopic model of pressureless collisionless dark matter. 
A more realistic particle-dark-matter halo with an inner edge, an outer cutoff, and dynamical matter perturbations would require a different construction and suitable matching conditions. 

There are several natural directions for future work. First, it would be important to extend the present construction to rotating black holes surrounded by Burkert-type halos, where the interplay between spin, superradiance, and halo structure may lead to a richer spectrum and possibly stronger observational signatures. Second, one may compare the Burkert profile systematically with other cored and cuspy halo models in order to identify which ringdown features are profile-dependent and which are more universal. Third, it would be worthwhile to go beyond the frozen-matter approximation adopted here and investigate whether perturbations of the halo itself can introduce additional effects in the spectrum or in the late-time tails. Finally, related observables such as polar gravitational modes, greybody factors, absorption properties, lensing, and shadow characteristics deserve further analysis within the same framework. We hope that the present results will serve as a useful step toward a more systematic understanding of effective environmental effects inspired by dark-matter density profiles in black hole perturbation theory and gravitational wave phenomenology.

\acknowledgements
This research was funded by the Guizhou Provincial Basic Research Program (Natural Science) Youth Guidance Program  (No.~QN [2025] 365), the National Natural Science Foundation of China (No.~12505064), the project of Young Scientific and Technical Talents Development of Education Department of Guizhou Province (No.~[2024] 79), the Guizhou Provincial Basic Research Program General Project (No.~MS [2026] 068), the Guizhou Provincial Basic Research Program (Natural Science) under Grant No.~QN [2025] 310. G. L and A. \"O. would like to acknowledge networking support of the COST Action CA21106 - COSMIC WISPers in the Dark Universe: Theory, astrophysics and experiments (CosmicWISPers), the COST Action CA22113 - Fundamental challenges in theoretical physics (THEORY-CHALLENGES), the COST Action CA21136 - Addressing observational tensions in cosmology with systematics and fundamental physics (CosmoVerse), the COST Action CA23130 - Bridging high and low energies in search of quantum gravity (BridgeQG), and the COST Action CA23115 - Relativistic Quantum Information (RQI) funded by COST (European Cooperation in Science and Technology). A. \"O. also thanks to EMU, TUBITAK, ULAKBIM (Turkiye) and SCOAP3 (Switzerland) for their support.

\appendix

\bibliography{DBH_ref}
\bibliographystyle{reference}
\end{document}